\documentclass[a4paper,fleqn,usenatbib]{mnras}

\usepackage[T1]{fontenc}
\usepackage{ae,aecompl}

\usepackage{graphicx}	% Including figure files
\usepackage{amsmath}	% Advanced maths commands
\usepackage{amssymb}	% Extra maths symbols
\usepackage{ulem}       %for revising. Remove before sending back
\usepackage{cancel}     %for revising. Remove before sending back
\usepackage{float}

\title[Eccentric TDE Disks in GRRMHD]{Global Simulations of Tidal Disruption Event Disk Formation via Stream Injection in GRRMHD}

\author[Brandon Curd]{
Brandon Curd$^{1}$\thanks{E-mail: brandon.curd@cfa.harvard.edu}
\\
$^{1}$ Harvard-Smithsonian Center for Astrophysics, 60 Garden Street, Cambridge, MA 02138, USA
}

\date{Accepted XXX. Received YYY; in original form ZZZ}

\pubyear{2021}

\begin{document}
\label{firstpage}
\pagerange{\pageref{firstpage}--\pageref{lastpage}}
\maketitle

% Abstract of the paper
\begin{abstract}
We use the general relativistic radiation magnetohydrodynamics code \verb=KORAL= to simulate the early stages of accretion disk formation resulting from the tidal disruption of a solar mass star around a super massive black hole (BH) of mass $10^6\,M_\odot$. We simulate the disruption of artificially more bound stars with orbital eccentricity $e\leq0.99$ (compared to the more realistic case of parabolic orbits with $e=1$) on close orbits with impact parameter $\beta\geq3$. We use a novel method of injecting the tidal stream into the domain. For two simulations, we choose $e=0.99$ and inject mass at a rate that is similar to realistic TDEs. We find that the disk only becomes mildly circularized with eccentricity $e\approx0.6$ within the $3.5$ days that we simulate. The rate of circularization is faster for pericenter radii that come closer to the BH. The emitted radiation is mildly super-Eddington with $L_{\rm{bol}}\approx3-5\,L_{\rm{Edd}}$ and the photosphere is highly asymmetric with the photosphere being significantly closer to the inner accretion disk for viewing angles near pericenter. We find that soft X-ray radiation with $T_{\rm{rad}} \approx 3-5\times 10^5$ K may be visible for chance viewing angles. Our simulations predict that TDEs should be radiatively inefficient with $\eta\approx0.009-0.014$. These are the first simulations which simultaneously capture the stream, disk formation, and emitted radiation.
\end{abstract}


%%%%%%%%%%%%%%%%%%%%%%%%%%%%%%%%%%%%%%%%%%%%%%%%%%

%%%%%%%%%%%%%%%%% BODY OF PAPER %%%%%%%%%%%%%%%%%%

\section{Introduction}

Stars orbiting a central black hole (BH) in a galactic nucleus can sometimes get perturbed such that their orbit brings them close enough to the BH to get tidally disrupted. Such events, which have been dubbed tidal disruption events (TDEs) or tidal disruption flares, result in a bright flare which peaks rapidly and is observable for years as it declines. The general theoretical understanding was developed decades ago \citep{Hills1975,Rees1988,Phinney1989,Evans1989}. The prediction was that a geometrically thick, circularized accretion disk will form with a density maximum near the tidal radius and will generate prompt emission in the optical and UV bands with a luminosity that decreases with time following a $t^{-5/3}$ power law. TDEs provide a rare glimpse into the nature of distant BHs which would ordinarily be quiescent and are thus expected to provide a laboratory for understanding BH physics. 

Since the initial discovery of TDEs with the X-ray telescope, \textit{ROSAT}, TDEs have been discovered in the X-ray, optical/UV, and radio (see \citealt{Komossa2015} for a review). The presence of outflows, possibly launched by an accretion disk, has been inferred in many cases due to radio emission \citep{Alexander2016,Alexander2017} and TDEs have also been observed to launch jets \citep{Bloom2011,Burrows2011,Zauderer2011,Cenko2012,Brown2015}. More recently, a handful of TDEs have been observed during the rise to peak \citep{Holoien2019,Holoien2020,Hinkle2021}. This bounty of observations is expected to grow significantly in the coming years, but the theoretical understanding of TDEs is still catching up in several respects.

On the theory side, the general understanding of the initial stellar disruption and stream evolution has been well developed \citep{Carter1982,Evans1989,Kochanek1994,Lodato2009,Brassart2010,Stone2013,Coughlin2015,Coughlin2016,Steinberg2019}. In addition, several authors have simulated the hydrodynamics of the disk formation \citep{Ramirez-Ruiz2009,Guillochon2013,Shiokawa2015,Bonnerot2016,Hayasaki2016,Liptai2019,Andalman2020,Bonnerot2020,Bonnerot2021}. These studies have demonstrated that the presence of a nozzle shock at pericenter as well as shocks due to the stream self interacting due to precession and the fallback of material towards the BH will lead to dissipation and disk formation. However, the numerical costs of global simulations have largely limited authors to studies of artificially more bound streams or TDEs around lower mass BHs.

The ultimate goal of theoretical studies is to understand the observed emission properties of TDEs. The emission is presumably linked to the properties of the disrupted stream and the BH, but the parameter space of TDEs is vast and requires precise scrutiny. Several authors have investigated the effect of the orbital parameters on the stream's binding energy distribution and the mass fall back rate. If the observed luminosity is strongly coupled to the mass fall back rate, it is expected that the rise to peak may allow for an independent determination of the pericenter radius, with stars disrupted on closer orbits having a sharper rise to peak. \citet{Liptai2019} show that the spin of the BH can also delay the peak of the luminosity. 

The possibility of determining the parameters of the disrupted star and the central BH from TDE observations was explored by \citet{Mockler2019}. However, without a complete library of TDE models it is difficult to break model degeneracies. Building such a library requires a precise modeling of the accretion flow properties. Numerically, this is complicated by the large time and distance scales involved. Additionally, evolving the radiation with the gas greatly increases computational overhead in global simulations. For this reason, much of the previous work to study TDEs through simulations has focused on the disruption, disk formation, and accretion flow separately.

The precise source of the observed radiation in TDEs has not been pinned down theoretically. \citet{Dai2018} proposed a unified model in which an inner accretion flow supplies X-rays which can be obscured depending on viewing angle. This possibility was also explored by \citet{Curd2019}. On the other hand, \citet{Piran2015} and \citet{Jiang2016} propose that the outflow from the stream self intersection can alone explain optically identified TDEs. Simulating the outflow and accretion disk together is necessary to directly discriminate relative contributions.

Radiation is also particularly important in super-Eddington flows as the gas in such cases is radiation dominated and the accretion disk may launch outflows with velocities in excess of $0.1-0.4c$ \citep{Sadowski2015,Sadowski2016b,Jiang2019}. The effects of radiation in the disk formation has only been studied by \citet{Bonnerot2021} thus far. They included realistic TDE parameters by using an injection of the outflow resulting from the stream self-intersection and found that the disk evolved towards a thin disk of nearly constant height rather than the thick geometry expected of a super-Eddington flow.

Attempts to simulate the resulting accretion flow in general relativistic radiation magnetohydronynamics (or GRRMHD, \citealt{Dai2018,Curd2019}) have demonstrated that if TDEs evolve towards a geometrically puffed up accretion disk as is expected in super-Eddington accretion flows, the emission will be X-ray dominated or optical/UV dominated depending on the viewing angle. However, while these simulations predict emission and outflows that are very similar to many observed TDEs, there is a significant uncertainty in the initial conditions and the photosphere radius/geometry in particular. \citet{Andalman2020} have demonstrated that even for very close stellar orbits the disk geometry is irregular and not likely to be significantly circularized even after several days; however, this has not been studied for more than a fraction of the fallback time for near parabolic disruptions. In addition, it is not entirely clear that the highly super-Eddington accretion rates assumed in \citet{Dai2018} and \citet{Curd2019} are applicable in most TDEs since these studies assumed circularization is highly efficient. As of this writing, it is still unclear if the nearly circularized disks that previous studies have found apply in real TDEs. In that regard, studying accretion flows forming in a more realistic manner is of significant interest.

Here we apply a new method of injecting the stream on its first return to pericenter in order to study the disk formation, emission, and photosphere geometry for a close disruption in a global simulation which accounts for the effects of radiation. We focus on the early stages of disk formation and emission. Simulating the radiative properties of close TDEs, although such events are expected to be less common, is beneficial to test our understanding of observed TDEs as they may make up an important part of the parameter space for TDEs. For example, \citet{Dai2015} demonstrated that TDEs for close orbits around lower mass BHs ($M_{\rm{BH}} < 5\times10^6\,M_\odot$) may be the population that produces soft X-ray TDEs. When the Large Synoptic Survey Telescope (LSST, \citealt{Ivezic2019}) comes on line, it is expected to observe 10 to 22 TDEs per night \citep{Bricman2020}. This would represent an unprecedented increase in the number of known TDEs and open up the opportunity to probe the statistics of TDEs. As such, improving the theoretical understanding of the emission properties of TDEs across the parameter space is prescient.

In this work, we consider the tidal disruption of a $1\,M_\odot$ star on a close, eccentric orbit around a BH of mass $10^6 \, M_\odot$.  We present GRRMHD simulations of the TDE disk formation using a novel method of injecting the stream into the simulation domain by defining the orbital parameters of the inflowing gas via TDE theory. Building on previous works, we expand the computational domain to capture the photosphere and measure the emerging luminosity for this class of TDEs. In addition, we for the first time study the effects of radiation on the evolution of the disk when both the incoming stream and the forming disk are present in the simulation domain. This has the benefit of allowing us to capture the photosphere geometry and expected emission throughout the evolution. We note that the incoming stream is artificially more bound in this work (i.e., we consider the stream to be on an elliptic trajectory with large eccentricity, rather than on a parabolic orbit), but we scale down the density of the incoming stream to values similar to those expected for near parabolic TDEs for four of the simulations that we discuss.

The paper is organized as follows. In \S\ref{sec:TDEphysics}, we provide a brief overview of the theoretical understanding of TDEs relevant for this work. In \S\ref{sec:nummethods}, we describe the numerical methods employed in the simulations and describe the treatment of radiation as well as the boundary conditions used to inject the stream. In \S\ref{sec:resultse99} and \S\ref{sec:resultse97}, we detail the results for each simulation presented herein. We discuss implications of these results in \S\ref{sec:discussion} and conclude in \S\ref{sec:conclusions}.

\section{Tidal Disruption Event Physics} \label{sec:TDEphysics}

Throughout this work, we use gravitational units to describe physical parameters. For distance we use the gravitational radius $r_g\equiv GM_{\rm{BH}}/c^2$ and for time we use the gravitational time $t_g\equiv GM_{\rm{BH}}/c^3$, where $M_{\rm{BH}}$ is the mass of the BH. Often, we set $G=c=1$, so the above relations would be equivalent to $r_g=t_g=M_{\rm BH}$. \footnote{For a BH mass of $10^6\,M_\odot$, the gravitational radius and time in CGS units are $r_g = 1.48\times 10^{11}$ cm and $t_g = 4.94$ s, respectively.} Occasionally, we restore $G$ and $c$ when we feel it helps to keep track of physical units.

We adopt the following definition for the Eddington mass accretion rate:
\begin{equation} \label{eq:mdotEdd}
  \dot{M}_{\rm{Edd}} = \dfrac{L_{\rm{Edd}}}{\eta_{\rm NT} c^2},
\end{equation}
where $L_{\rm{Edd}} = 1.25\times 10^{38}\, (M/M_\odot)\, {\rm erg\,s^{-1}}$ is the Eddington luminosity, $\eta_{\rm{NT}}$ is the radiative efficiency of a thin disk around a BH with spin parameter $a_*$ (which is often referred to as the Novikov-Thorne efficiency),
\begin{equation} \label{eq:etaNT}
  \eta_{\rm{NT}} = 1 - \sqrt{1 - \dfrac{2}{3 r_{\rm{ISCO}}}},
\end{equation}
and $r_{\rm{ISCO}}=3+Z_2 - \sqrt{(3-Z_1)(3+Z_1+2Z_2)}$ is the radius of the Innermost Stable Circular Orbit (ISCO, \citealt{Novikov1973}) in the Kerr metric, where $Z_1 = 1 + (1-a_*^2)^{1/3}\left((1+a_*)^{1/3}+(1-a_*)^{1/3}\right)$ and $Z_2 = \sqrt{3a_*^2 + Z_1^2}$. For $a_* =0$, the efficiency is $\eta_{\rm{NT}}=0.05712$.

A star which has been captured by a SMBH will be disrupted when it can no longer be held together by its self-gravity. This occurs at radii less than the tidal radius,
\begin{equation} \label{eq:eq1}
  R_t/r_g = 47 m_6^{-2/3} m_*^{-1/3}r_*,
\end{equation}
where $m_6=M_{\rm{BH}}/10^6\,M_\odot$ is the mass of the SMBH, $m_*=M_{\rm{*}}/M_\odot$ is the mass of the disrupted star, and $r_*=R_{\rm{*}}/R_\odot$ is its radius. It is common to describe the disruption in terms of the impact parameter, $\beta$, which is defined as the ratio between the tidal radius and pericenter separation such that $\beta \equiv R_t/R_p$. A full disruption occurs for $\beta \geq 1$.

If hydrodynamical forces are neglected, then the change in the specific binding energy of the fluid in the star as a result of the tidal interaction can greatly exceed the internal binding energy of the star \citep{Rees1988}. As a result, a spread in binding energy is imparted on the stellar material. \citet{Stone2013} find that the spread in orbital energy $\Delta\epsilon$ is insensitive to $\beta$ since the energy is essentially frozen in at the tidal radius. This spread is then given by:
\begin{equation} \label{eq:eq4}
  \Delta\epsilon \approx 4.3\times10^{-4} \dfrac{m_6^{1/3}m_*^{2/3}}{r_*}c^2.
\end{equation} 
The orbital binding energy of the most/least bound material is given by $\epsilon_{\rm{mb}} = \epsilon_* - \Delta\epsilon/2$ and $\epsilon_{\rm{lb}} = \epsilon_* + \Delta\epsilon/2$. Here $\epsilon_*$ is the initial orbital binding energy of the star. For parabolic orbits, which have $\epsilon_*=0$, the spread in binding energy leads to half of the mass remaining bound and the other half being ejected. However, if the star is on an elliptical orbit (gravitationally bound to the SMBH) and has an initial binding energy $\epsilon_* < -\Delta\epsilon/2$, then all of the stellar material remains bound after disruption and returns to pericenter in a finite time. 

In this work, we study the tidal stream of a $1M_\odot$ main sequence star around a $10^6M_\odot$ SMBH for elliptical ($e<1$), close ($\beta>1$) orbits. This leads to a disruption where $\epsilon_*=-\beta c^2 (1-e)/2(R_t/r_g) < -\Delta\epsilon/2$. The orbit of the disrupted star is assumed to be aligned with the equatorial plane of the BH spin vector. The orbital period of the most bound material is given by $t_{\rm{mb}}=2\pi (-2\epsilon_{\rm{mb}})^{-3/2}$ and that of the least bound material by $t_{\rm{lb}}=2\pi (-2\epsilon_{\rm{lb}})^{-3/2}$. Thus there is a difference in the arrival times of the most and least bound material of $\Delta t = t_{\rm{lb}} - t_{\rm{mb}}$. The commonly used 'fallback time' is the time it takes for the most bound material to return to pericenter following disruption; therefore, we set the fallback time to $t_{\rm{fallback}}=t_{\rm{mb}}$.

As it makes its first pericenter passage, the stream precesses due to relativistic effects. We adopt a similar method to \citet{Dai2015} to quantify this precession. On its first pericenter passage, the precession angle may be approximated by
\begin{equation} \label{eq:precessionangle}
    \Delta\phi = \dfrac{6\pi}{a(1-e^2)}.
\end{equation}
Note that we have expressed $\Delta\phi$ using gravitational units so the semi-major axis $a$ is given in gravitational radii. Treating the orbits of the incoming stream that has yet to pass through pericenter and the already precessed stream as ellipses, the self intersection between the incoming material and material that has precessed occurs at the radius
\begin{equation} \label{eq:selfintersection}
  R_{\rm{SI}} = \dfrac{(1+e)R_t}{\beta(1-e \cos (\Delta\phi/2))}.
\end{equation}
The initial evolution of the disk is expected to be driven by dissipation of kinetic energy at this point. As the velocity of the stream elements is greater at smaller radii, the rate of dissipation will also be greater for closer orbits (larger $\beta$).

%%%%%%%%%%%%%%%%% Table 1
\begin{table*}
  \centering
  \begin{tabular}{l c c c c c c}
    \hline
    \hline
     & \texttt{e99\_b5\_01} & \texttt{e99\_b3\_01}  & \texttt{e97\_b5\_01}  & \texttt{e97\_b5\_02}   & \texttt{e97\_b5\_03}  & \texttt{e97\_b5\_04}\\
    \hline
    \textbf{magnetic field?} & no & no & yes & no & no & no \\
    \textbf{radiation?} & yes & yes & yes & no &yes & no \\
    $e$  & $0.99$ & $0.99$ & $0.97$ & $0.97$ & $0.97$ & $0.97$ \\
    $\beta$  & $5$ & $3$ & $5$ & $5$ & $5$ & $5$ \\
    $M_{\rm{inj,tot}}$ & $0.013\,M_\odot$  & $0.016\,M_\odot$ &  $1\,M_\odot$ &  $1\,M_\odot$ &  $0.04\,M_\odot$ &  $0.04\,M_\odot$ \\
    $\dot{M}_{\rm{0}} (\dot{M}_{\rm{Edd}})$ & 133 & 133 & 19,330 & 19,330  & 800 & 800 \\ 
    $a_*$  & $0$ & $0$ & $0.9$ & $0.9$ & $0$ & $0$ \\
    $N_r \times N_\theta \times N_\phi$ & $160\times 128 \times 128$ & $160\times 128 \times 128$ & $160\times 128 \times 128$  & $160\times 128 \times 128$ & $96\times 96 \times 128$  & $96\times 96 \times 128$ \\
    $R_{\rm{inj}}\,(r_g)$  & 200 & 400 & 500 & 500 & 200 & 200 \\ 
    $R_{\rm{SI}}\,(r_g)$  & 40 & 168 & 40 & 40 & 40 & 40 \\ 
    $R_{\rm{min}} \,(r_g) /R_{\rm{max}} \,(r_g)$ & $1.8/5\times10^4$ & $1.8/5\times10^4$ & $1.3/10^5$ & $1.3/10^5$  & $1.8/10^3$  & $1.8/10^3$ \\
    $t_{\rm{fallback}}\,(t_g)$ & 108,825 & 179,946 & 28,830 & 28,830 & 28,830 & 28,830 \\
    $\Delta t_{\rm{inj}}\,(t_g)$ & 284,925 & 1,910,084 & 14,467 & 14,467 & 14,467 & 14,467 \\
    $t_{\rm{max}}\,(t_g)$ & 60,000 & 60,000 & 40,000 & 60,000 & 60,000 & 60,000 \\
    \hline
  \end{tabular}
  \caption{Simulation parameters and properties of the six simulations. We specify whether the magnetic field and radiation were evolved with the gas. We also specify the eccentricity ($e$), impact parameter ($\beta$), total mass injected during the injection phase ($M_{\rm{inj,tot}}$), the peak injection rate of mass into the domain ($\dot{M}_0$), the spin of the BH ($a_*$), the radius at which mass is injected ($R_{\rm{inj}}$), the self-intersection radius of the stream ($R_{\rm{SI}}$), the inner and outer radial boundaries of the simulation box, the  resolution of the grid, the fallback time of the stream ($t_{\rm{fallback}}$), the time at which the least bound material is injected ($\Delta t_{\rm{inj}}$), and the total run time for each simulation ($t_{\rm{max}}$).}
  \label{tab:tab1}
\end{table*}
%%%%%%%%%%%%%%%% End Table 1

\section{Numerical Methods} \label{sec:nummethods}
The simulations presented in this work were performed using the general relativistic radiation magnetohydrodynamical (GRRMHD) code \verb=KORAL= \citep{Sadowski2013,Sadowski2014,Sadowski2017} which solves the conservation equations in a fixed, arbitrary spacetime using finite-difference methods. We solve the following conservation equations:
\begin{align}
  (\rho u^\mu)_{;\mu} &= 0, \label{eq:eq5} \\
  (T^\mu_\nu)_{;\mu} &= G_\nu, \label{eq:eq6} \\
  (R^\mu_\nu)_{;\mu} &= -G_\nu, \label{eq:eq7} 
\end{align}
where $\rho$ is the gas density in the comoving fluid frame, $u^\mu$ are the components of the gas four-velocity as measured in the ``lab frame'', $T^\mu_\nu$ is the MHD stress-energy tensor in the ``lab frame'',
\begin{equation} \label{eq:eq9}
  T^\mu_\nu = (\rho + u_g+ p_g + b^2)u^\mu u_\nu + (p_g + \dfrac{1}{2}b^2)\delta^\mu_\nu - b^\mu b_\nu,
\end{equation}
$R^\mu_\nu$ is the stress-energy tensor of radiation, and $G_\nu$ is the radiative four-force which describes the interaction between gas and radiation \citep{Sadowski2014}. Here $u_g$ and $p_g=(\gamma - 1)u_g$ are the internal energy and pressure of the gas in the comoving frame and $b^\mu$ is the magnetic field four-vector which is evolved following the ideal MHD induction equation \citep{Gammie2003}. 

The radiative stress-energy tensor is obtained from the evolved radiative primitives, i.e. the radiative rest-frame energy density and its four velocity following the M1 closure scheme modified by the addition of radiative viscosity \citep{Sadowski2013,Sadowski2015}.

The interaction between gas and radiation is described by the radiation four-force $G_\nu$. The opposite signs of this quantity in the conservation equations for gas and radiation stress-energy (equations \ref{eq:eq6}, \ref{eq:eq7}) reflect the fact that the gas-radiation interaction is conservative, i.e. energy and momentum are transferred between gas and radiation. For a detailed description of the four-force see \citet{Sadowski2017}. We include the effects of absorption and emission via the electron scattering opacity ($\kappa_{\rm{es}}$) and free-free asborption opacity ($\kappa_{\rm{a}}$) and assume a Solar metal abundance for the gas. 

We use modified Kerr-Schild coordinates with the inner edge of the domain inside the BH horizon. The radial grid cells are spaced logarithmically in radius and the cells in polar angle $\theta$ have smaller widths towards the equatorial plane. The cells are equally spaced in azimuth. At the inner radial boundary ($R_{\rm{min}}$), we use an outflow condition while at the outer boundary ($R_{\rm{max}}$) we use a similar boundary condition and in addition prevent the inflow of gas and radiation. At the polar boundaries, we use a reflective boundary. We exclude a small region of the polar angle such that $\theta_{\rm{min}}=0.005\pi$ and $\theta_{\rm{max}}=0.995\pi$ to reduce computation time near the horizon where the limit on the time step in the polar azimuthal directions becomes small. We employ a periodic boundary condition in azimuth and the grid covers $-\pi \leq \phi \leq \pi$.

We quantify the resolution of the fastest growing mode of the magnetorotational instability (MRI, \citealt{Balbus1991}) by computing the quantities:
\begin{align}
  Q_\theta = \dfrac{2\pi}{\Omega dx^\theta}\dfrac{|b^\theta|}{\sqrt{4\pi\rho}}, \label{eq:eq10} \\
  Q_\phi = \dfrac{2\pi}{\Omega dx^\phi}\dfrac{|b^\phi|}{\sqrt{4\pi\rho}}, \label{eq:eq11}
\end{align}
where $dx^i$ (the grid cell) and $b^i$ (the magnetic field) are both evaluated in the orthonormal frame, $\Omega$ is the angular velocity, and $\rho$ is the gas density. Numerical studies of the MRI have shown that values of $Q_\theta$ and $Q_\phi$ in excess of at least 10 are needed to resolve the fastest growing mode \citep{Hawley2011}. We discuss later how the MRI evolves in the one simulation in which we include a magnetic field.

\subsection{Injection of TDE Stream}

Previous hydrodynamical simulations of TDE disks have been performed by starting with smooth particle hydrodynamics simulations of the disruption to obtain the initial data. In the present work, we inject the TDE stream at an interior boundary using a description of the fluid based on TDE theory. The primary motivation for this approach is the possibility of studying a broad range of TDE disks by simply changing the properties of the injected stream. This is the first work in which we employ this numerical setup.

The simulation domain is initialized with a low density atmosphere with a density profile that scales with $r^{-2}$. The atmosphere is initialized with a constant radiation temperature of $T_{\rm{atm}}=10^5$ K. We inject the TDE stream at an interior boundary (which is henceforth referred to as the `injection boundary') $R_{\rm{min}} < R_{\rm{inj}} < R_{\rm{max}}$. The injection boundary radii used in each simulation are shown in Table \ref{tab:tab1}.

The mass inflow rate at the injection boundary decreases over time following a $\dot{M}\propto t^{-5/3}$ profile. The exact description of the mass injection is given by:

\begin{equation}
    \dot{M}_{\rm{inj}}(t) = \dot{M}_0 \left(\dfrac{t}{t_{\rm{fallback}}} + 1\right)^{-5/3}
\end{equation}
where $\dot{M}_0$ is the peak mass inflow rate, and $t$ is the time since the beginning of injection in gravitational units. Since the disrupted stellar material for the two eccentricities we consider, $e=0.97$ and $e=0.99$, returns in a finite time, we turn off the mass injection after $t > \Delta t_{\rm{inj}} = (t_{\rm{lb}} - t_{\rm{mb}})$. We list the time at which the least bound material is injected for each simulation in Table \ref{tab:tab1}. Note that in our set up, the most bound material is injected at the simulation time $t=0$. After the least bound material is injected at $t=\Delta t_{\rm{inj}}$, we switch to a reflecting boundary condition for the cells at the injection boundary. All the simulations are run for a total run time of $t_{\rm max} = 60,000M$ with the exception of \texttt{e99\_b5\_01} which has a run time of $t_{\rm max} = 40,000M$. In each case, this duration is longer than $\Delta t_{\rm{inj}}$ for $e=0.97$ and so stream injection ceases part-way through these simulations. However, for $e=0.99$, $t_{\rm max}$ is less than $\Delta t_{\rm{inj}}$, and for these simulations stream injection continues up to the end of these simulations.

The trajectory of the incoming fluid is determined by its specific binding energy and angular momentum. The angular momentum is fixed to the value corresponding to the pericenter radius of the TDE stream $l=\sqrt{2R_{\rm{p}}}$. The radial velocity is determined from the specific binding energy, which varies as:
\begin{equation}
    \epsilon_{\rm{inj}}(t)=\epsilon_0 \left(\dfrac{t}{t_{\rm{fallback}}}+1\right)^{-2/3}.
\end{equation}
The radial velocity is then set by:
\begin{equation}
    v_{\rm{inj}}(t) = -\sqrt{\dfrac{2}{R_{\rm{inj}}} + 2\epsilon_{\rm{inj}}(t)} ~.
\end{equation}
These estimates are all based on a Newtonian approximation, which is sufficiently accurate for our purpose.

The maximum stream thickness at pericenter can be estimated to be of order $(H/R)_{\rm{max}} = R_\odot/R_p \sim 0.01$. Because of resolution limitations, especially in the azimuthal $\phi$ direction, we choose to inject gas with a larger scale height of $H/R=0.05$, which still covers only two cells in azimuth. We inject the gas with a constant mass density since the resolution at the injection point is too poor to include a density profile.

The gas temperature is set to $T_{\rm{inj}}=10^5$ K at the injection boundary. This temperature is used to set the total pressure of the stream. For simulations where radiation is included, we use the initial gas pressure obtained from $T_{\rm{inj}}$ to split the internal energy into gas and radiation energy density by solving the condition $p_{\rm{tot}} = p_{\rm{gas}}+p_{\rm{rad}}$ and finding a new gas and radiation temperature which assumes thermal equilibrium of the gas.

For the simulation where we include a magnetic field (\texttt{e97\_b5\_01}), we inject a magnetic field with a poloidal field geometry and a magnetic pressure ratio of $\beta_{\rm{mag}} \equiv p_{\rm{mag}}/(p_{\rm{gas}}+p_{\rm{rad}}) = 0.01$. This choice of $\beta_{\rm{mag}}$ guarantees that the magnetic field does not impact the gas dynamics as the disk forms.

\subsection{Simulation Details}

We list the simulations presented in this work in Table \ref{tab:tab1}. The name of each simulation is listed in the top row and each name describes the eccentricity and impact parameter defining the binding energy and angular momentum of the incoming material. We also add a numerical tag at the end to differentiate simulations with similar disruption parameters. For example, \texttt{`e99\_b5\_01'} was initialized with $e=0.99$ and $\beta=5$, which is also indicated in Table \ref{tab:tab1}.

In this study, we perform simulations in which the binding energy and angular momentum of the stream are set assuming the disruption of a Sun-like star. For four of the simulations (\texttt{e99\_b5\_01}, \texttt{e99\_b3\_01}, \texttt{e97\_b5\_03}, and \texttt{e97\_b5\_04} in Table \ref{tab:tab1}), we artificially scale down the mass injection rate. For \texttt{e99\_b5\_01} and \texttt{e99\_b3\_01}, the peak mass injection rate is approximately that expected for a parabolic TDE. Note that the binding energy and angular momentum for a $1\, M_\odot$ star are maintained in spite of this modification to the injected stream. For \texttt{e97\_b5\_01} and \texttt{e97\_b5\_02}, we inject a full solar mass over the course of the mass injection. As we discuss in \S\ref{sec:resultse97}, radiation from the shocks and forming disk is able to diffuse out and push on the gas rather than merely being advected if the density of the incoming stream is lowered to more realistic values. Since we wish to study the impact of radiation in the early evolution of TDE disks, we choose to artificially decrease the stream density in these cases. We discuss the consequences of this in \S\ref{sec:discussion}.

%%%%
% Beginning of section 4
%%%%
\section{Near Parabolic Simulations} \label{sec:resultse99}

We first discuss the simulations \texttt{e99\_b5\_01} and  \texttt{e99\_b3\_01}. Although not quite parabolic, the injected streams in these models are close enough to parabolic to be reasonable representations of real TDEs. We focus our discussion on \texttt{e99\_b5\_01}, but we contrast results with \texttt{e99\_b3\_01} where key differences arise.

%%%%%%%%%%
% Begin Figure
%%%%%%%%%
\begin{figure}
    \centering{}
	\includegraphics[width=\columnwidth]{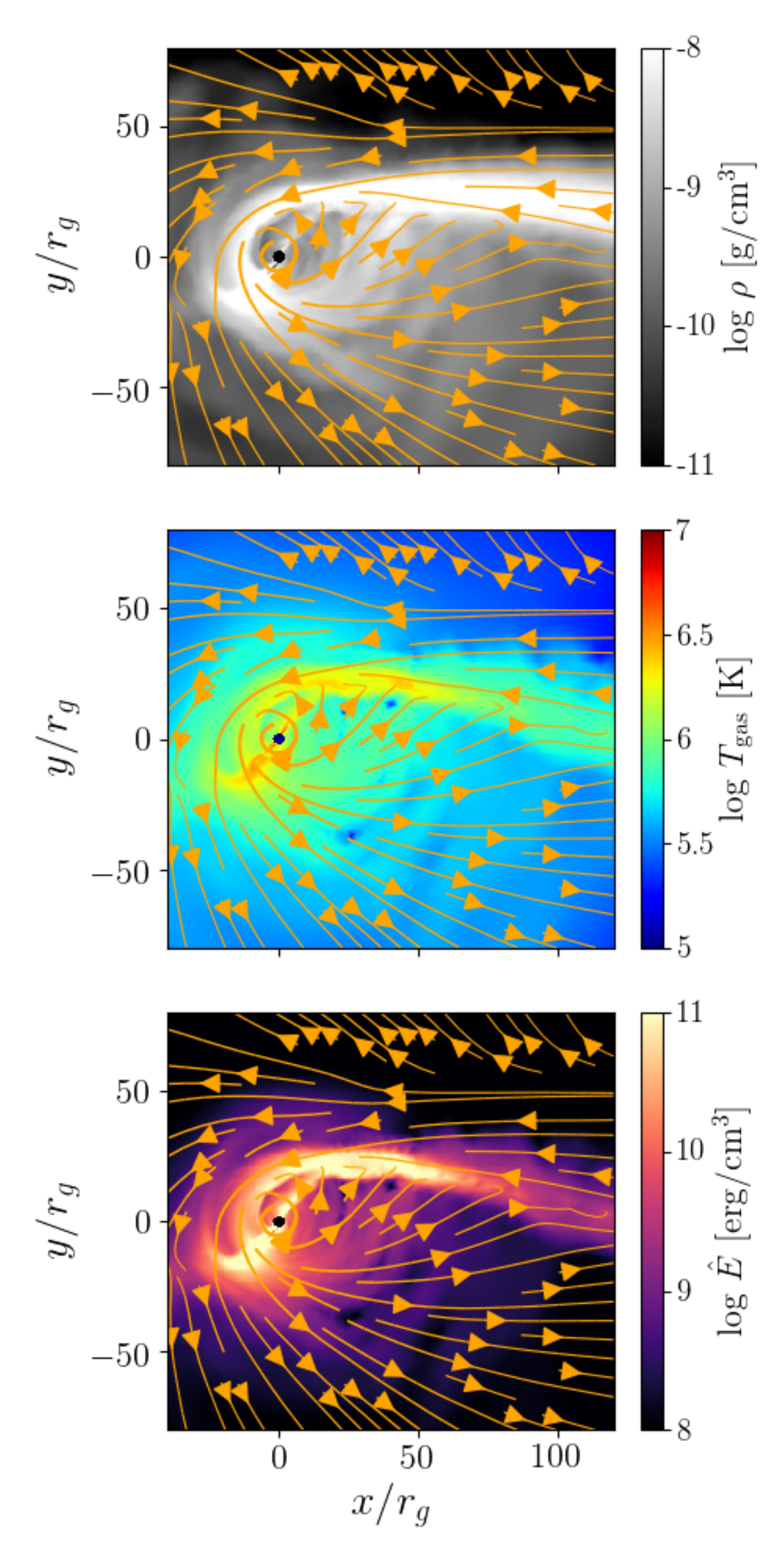}
    \caption{Here we show an equatorial slice of the self intersection region for \texttt{e99\_b5\_01} at $t=3,000 \,t_g$. The colors show the gas density (top), gas temperature (middle) and radiation energy density (bottom). In each panel, the orange arrows show the velocity vectors. The self intersection occurs over a range of radii from $10-100\, r_g$. The shock is indicated by an increase in temperature and radiation energy density. There is also seemingly a secondary shock near pericenter as indicated by the increase in temperature and radiation energy density to the left of the BH.}
    \label{fig:e99beta5hm}
\end{figure}
%%%%%%%%%%
% End Figure
%%%%%%%%%

%%%%%%%%%%
% Begin Figure
%%%%%%%%%
\begin{figure*}
    \centering{}
	\includegraphics[width=\textwidth]{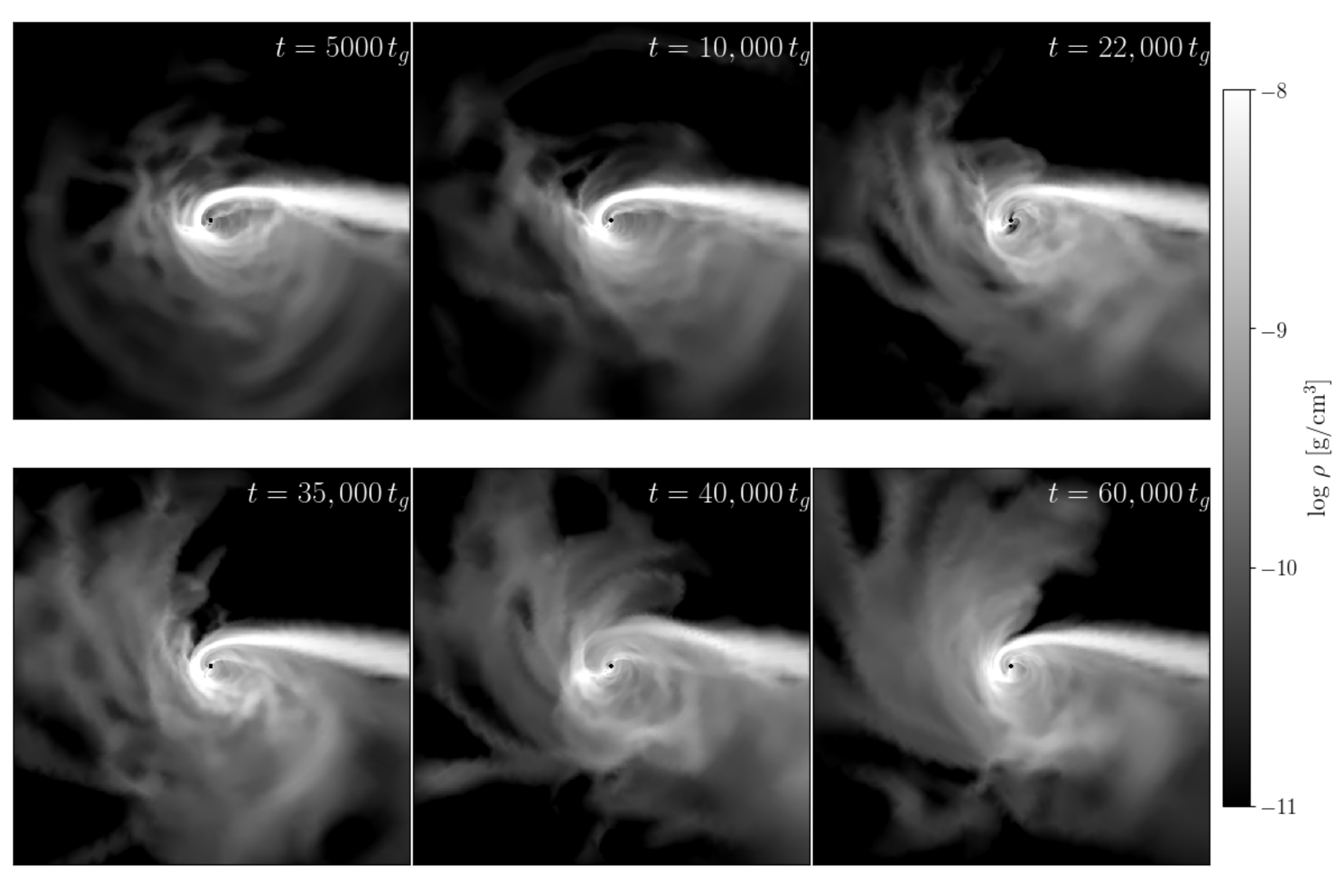}\\
    \caption{Here we show selected snapshots of the gas density (colors) for a slice through the mid plane ($\theta=\pi/2$) for \texttt{e99\_b5\_01} to highlight parts of the evolution. The scale of each image is $400r_g\times400r_g$ and the BH is centered in the image. There are multiple events in the evolution, similar to that shown in the bottom middle panel, where the incoming stream is fully disrupted.}
    \label{fig:e99beta5evol}
\end{figure*}

\begin{figure}
    \centering{}
	\includegraphics[width=\columnwidth]{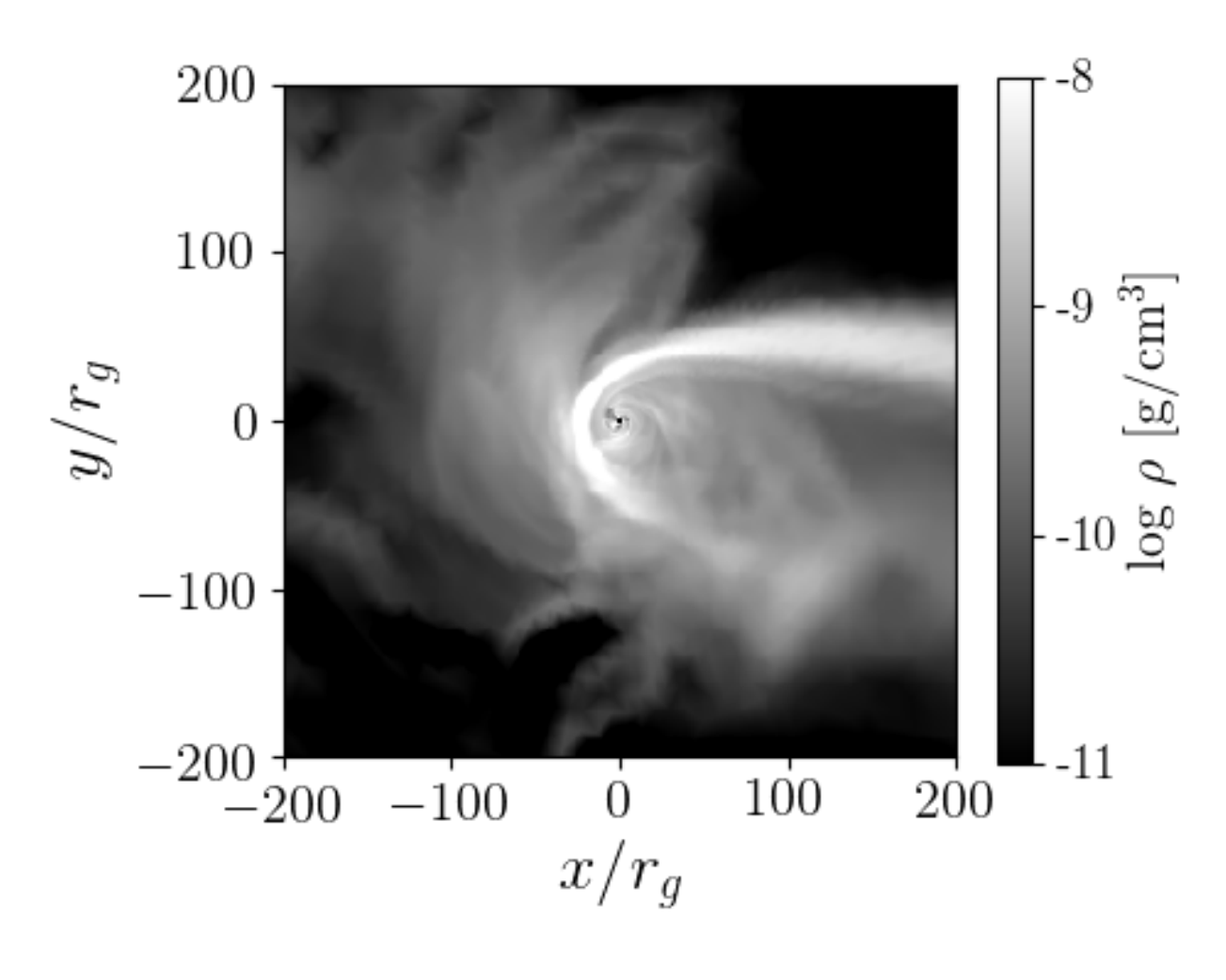}\\
    \caption{Here we a snapshot of the gas density (colors) for a slice through the mid plane ($\theta=\pi/2$) for \texttt{e99\_b3\_01} at $t=60,000\,t_g$. The scale of each image is $400r_g\times400r_g$ and the BH is centered in the image.}
    \label{fig:e99b3thsli}
\end{figure}
%%%%%%%%%%
% End Figure
%%%%%%%%%

\subsection{Dynamics}

As the incoming gas passes through pericenter, it undergoes relativistic orbital precession and collides and shocks with the incoming stream. The energy available for dissipation in the interaction is determined by the radius of self-intersection. For the $\beta=5$ model, the theoretical self intersection radius is $R_{\rm{SI}}\approx 40$, though we note that the stream appears to spread out after the nozzle shock which results in a range of radii for self intersection between $10-100\,r_g$. The typical collision velocity at the self-intersection point is $v \approx 0.2 c$. For the $\beta=3$ model, the self intersection radius is $R_{\rm{SI}}\approx 168$ and the typical collision velocity is slightly lower at $v \approx 0.1 c$.

We note that the stream has a significant radial width as it passes through pericenter, which is characteristic of eccentric streams. The fluid elements that orbit closer to the BH precess more than those farther away. As a result, the gas appears to fan out as it passes through the nozzle. This effect, which can be seen in Figure \ref{fig:e99beta5hm}, leads to the gas that collides with the returning stream having lower density. This is not expected in realistic TDEs (Bonnerot, Private Communication).

Due to vertical crossing at pericenter, the gas forms a nozzle region but this feature is poorly resolved in the present simulations given our choice of grid. Similar to \citet{Sadowski2016a}, we find that this region is not as narrow as in parabolic disruptions in part due to the eccentricity of the treated disruption, but the vertical extent may also be artificially larger due to the artificial stream thickness that we employ. There is some dissipation in the nozzle, which can be seen in the increase in temperature near pericenter (Figure \ref{fig:e99beta5hm}) but the nozzle region in our simulations is only marginally resolved (we have 10 cells in the nozzle region). The qualitative results are not expected to be impacted by this.

The dissipation of kinetic energy in the self-intersection shock leads to significant heating of the gas, with the inner accretion disk reaching temperatures $T_{\rm{gas}}\approx10^6$ K. This in turn causes a strong outflow of gas. A significant fraction of the shocked gas becomes more bound and falls directly into the BH. An even larger fraction becomes unbound and gets ejected in an outflow carrying a significant amount of kinetic energy. As discussed in \cite{Lu2020}, the fraction of gas that becomes unbound due to shocks is sensitive to both the stellar parameters and the BH mass, and the maximum expected fraction of gas that becomes unbound is 50\%. We discuss the outflows in our models in more detail in \S\ref{sec:e99outflows}.

We show the full evolution of the forming disk for \texttt{e99\_b5\_01} in Figure \ref{fig:e99beta5evol}. We find that there are various epochs during the evolution of this model (true also for \texttt{e99\_b3\_01)}, such as the $t=40,000t_g$ epoch shown in the bottom middle panel in Figure \ref{fig:e99beta5evol}, where the incoming stream is temporarily disrupted due to the violent self intersection. These events are accompanied by significant shock heating and gas being flung onto a wide range of orbits. Such events were also found in a TDE simulation of a star on a close orbit by \citet{Andalman2020}. We discuss the properties of the disruptions in our simulations in more detail in \S\ref{sec:discussion}. We also show the final state of the disk in the mid plane for \texttt{e99\_b3\_01} in Figure \ref{fig:e99b3thsli}.

\begin{figure}
    \centering{}
	\includegraphics[width=\columnwidth]{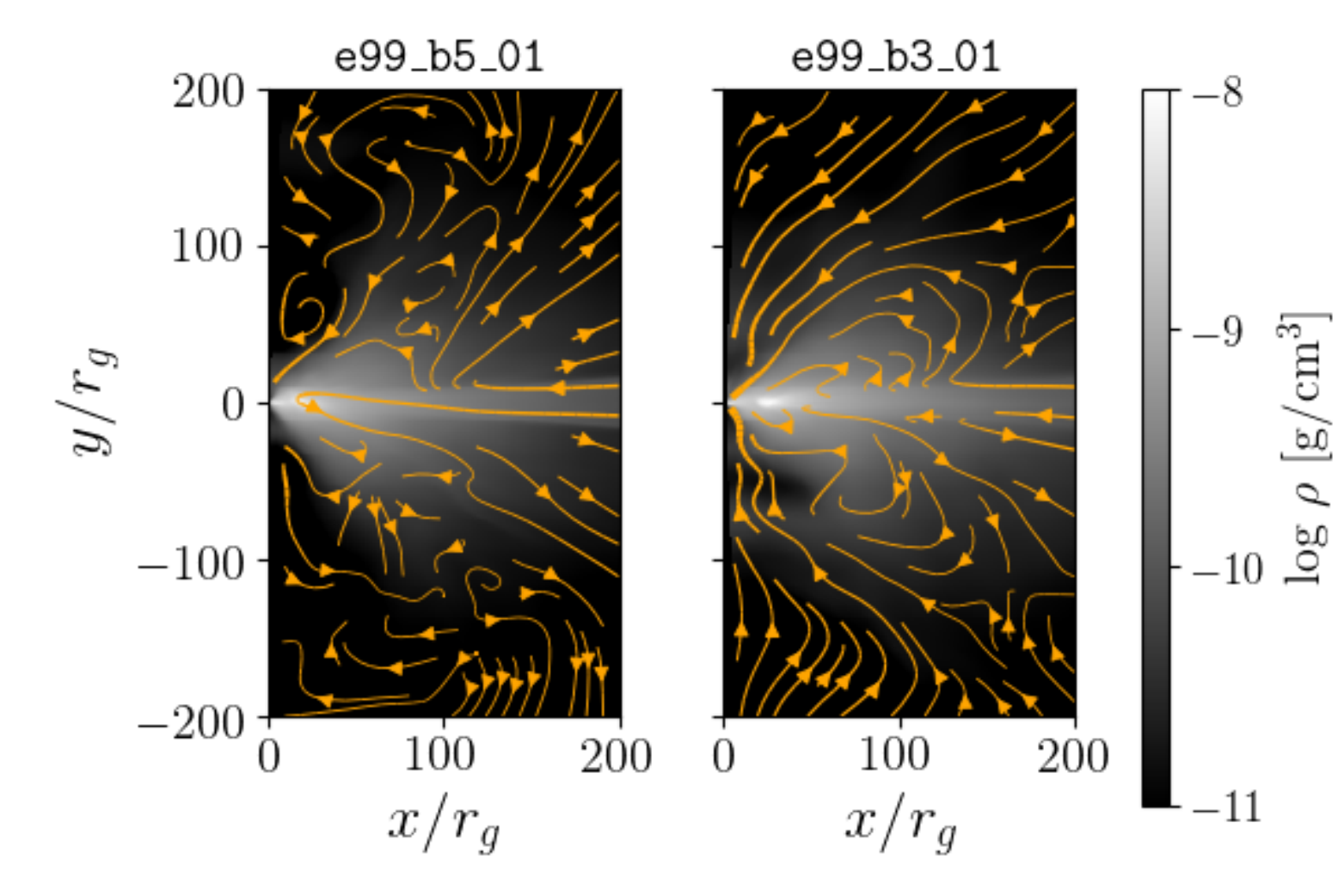}\\
    \caption{Here we show time and azimuth averages of the gas density (colors) and fluid velocity (orange arrows) for \texttt{e99\_b5\_01} (left) and \texttt{e99\_b3\_01} (right). Each figure is averaged over $59,000-60,000\, t_g$. The disks are of similar thickness, and the density maximum of the disk is near the pericenter radius since the stream still passes through the disk. Interestingly, \texttt{e99\_b5\_01} appears to have an outflow near the poles while for \texttt{e99\_b3\_01} material is falling inwards near the poles.}
    \label{fig:e99phiavg}
\end{figure}

We show the vertical structure of the disk at the end of the simulations \texttt{e99\_b5\_01} and \texttt{e99\_b3\_01} in Figure \ref{fig:e99phiavg}. The resulting disk is puffed up with the density maximum occuring between the pericenter radius and the circularization radius ($R_{\rm{circ}}\equiv2R_p$). The entire outflow and disk is radiation pressure dominated, with the disk reaching a pressure ratio $\beta_{\rm{rad}}\equiv p_{\rm{rad}}/p_{\rm{gas}} \approx 10^5$ near the density maximum. 

\subsection{Accretion Disk Properties}

%%%%%%%%%%
% Begin Figure
%%%%%%%%%
%%%%%%%%%%
% End Figure
%%%%%%%%%

%%%%%%%%%%
% Begin Figure
%%%%%%%%%
\begin{figure}
    \centering{}
	\includegraphics[width=\columnwidth]{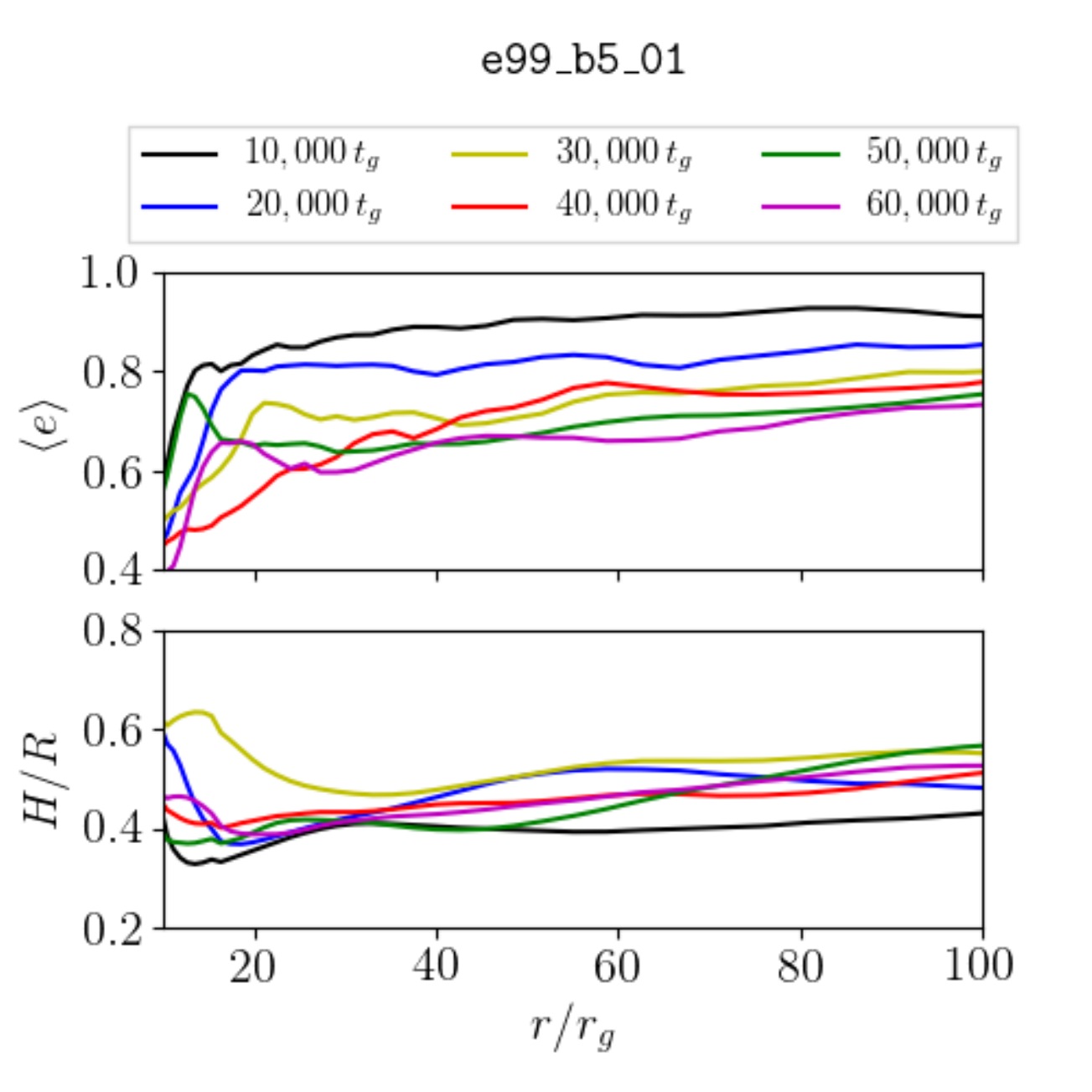}\\
    \caption{Here we show the mass weighted eccentricity (top) and density scale height of the disk (bottom) at 6 epochs for \texttt{e99\_b5\_01}.}
    \label{fig:e99b5ecc}
\end{figure}
%%%%%%%%%%
% End Figure
%%%%%%%%%

%%%%%%%%%%
% Begin Figure
%%%%%%%%%
\begin{figure}
    \centering{}
	\includegraphics[width=\columnwidth]{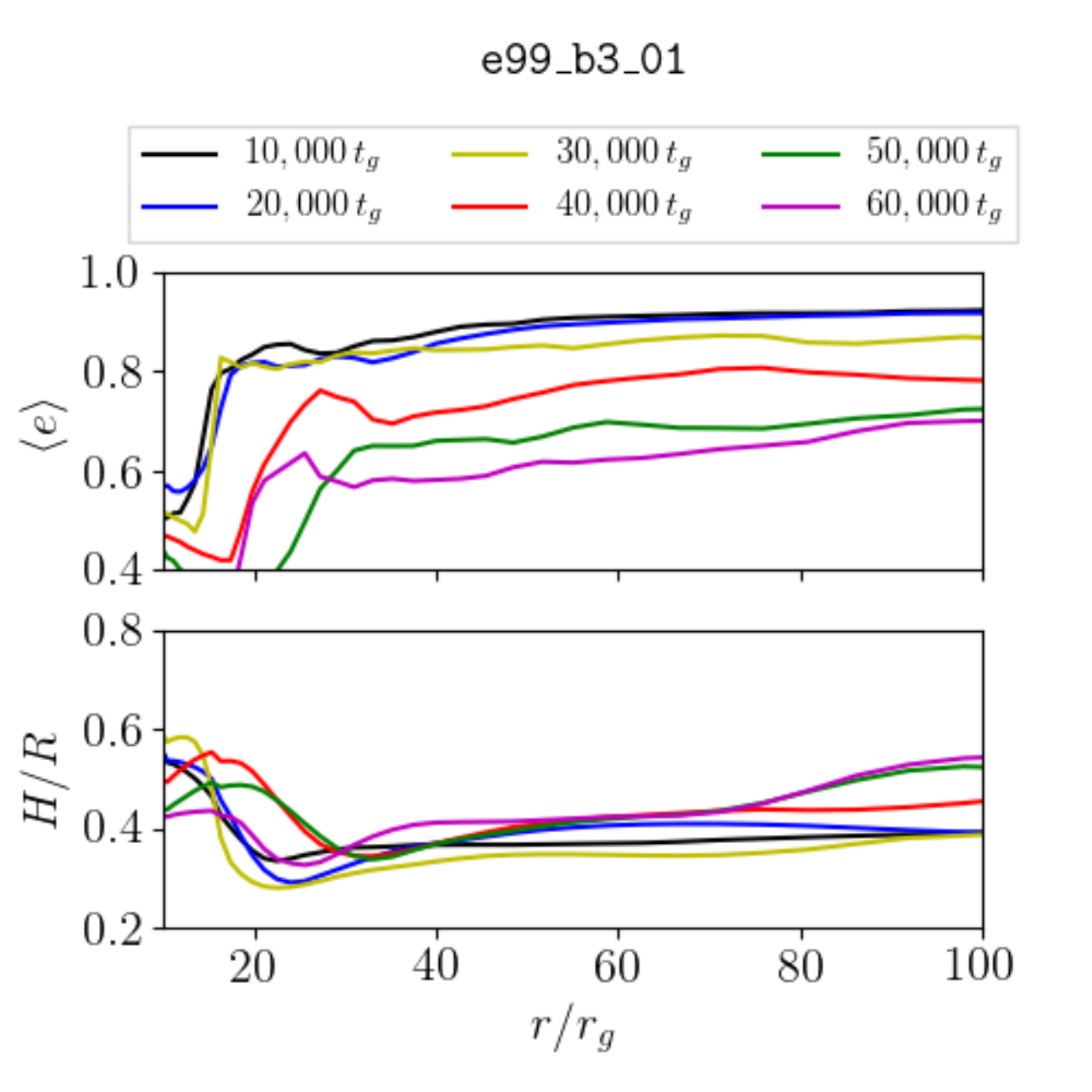}\\
    \caption{The same as Figure \ref{fig:e99b5ecc} but for \texttt{e99\_b3\_01}.}
    \label{fig:e99b3ecc}
\end{figure}
%%%%%%%%%%
% End Figure
%%%%%%%%%

%%%%%%%%%%
% Begin Figure
%%%%%%%%%
\begin{figure}
    \centering{}
	\includegraphics[width=\columnwidth]{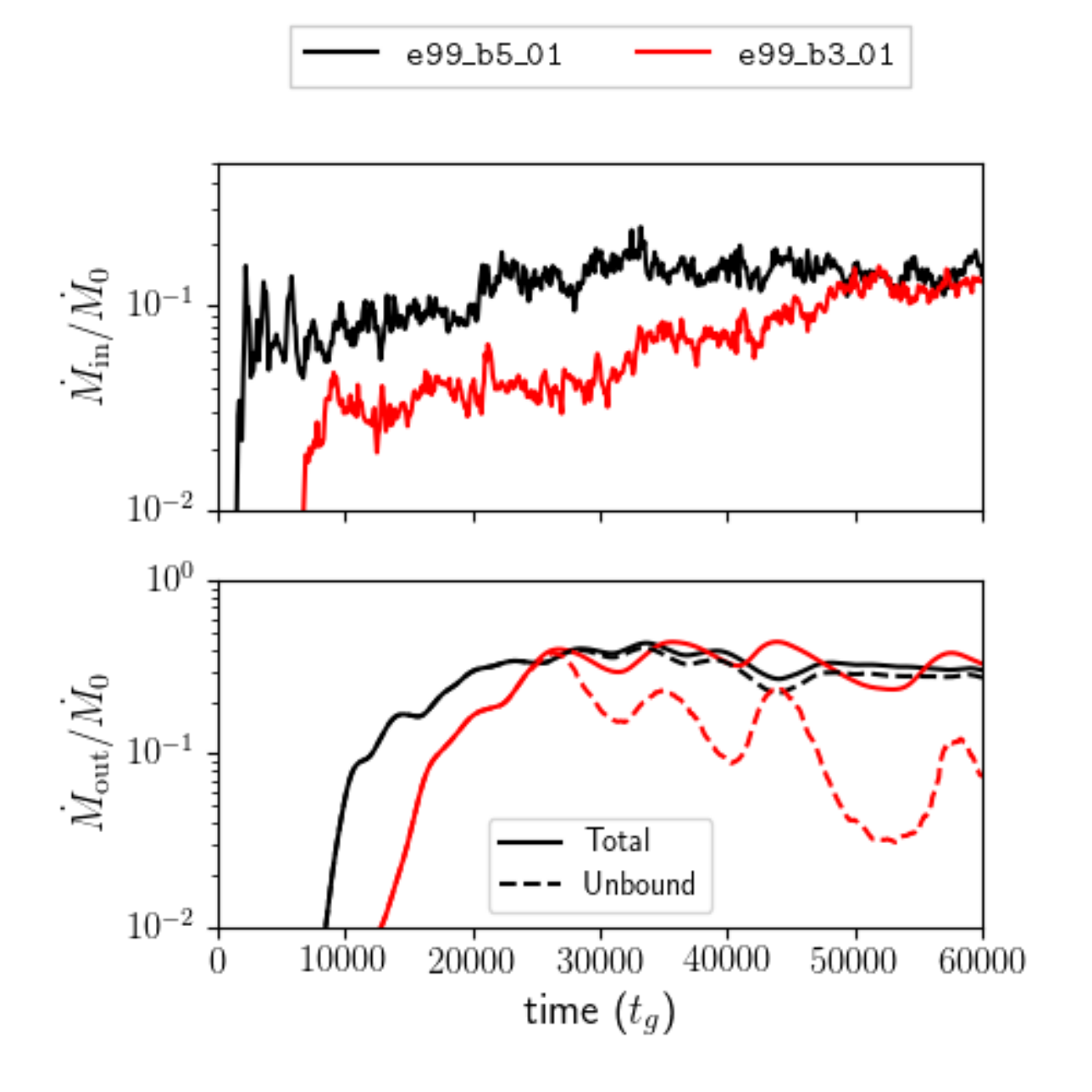}\\
    \caption{Here we show the inflow rate of mass crossing the horizon (top) and the outflow rate of mass crossing $r=1000\,r_g$ (bottom) for \texttt{e99\_b5\_01} (black lines) and  \texttt{e99\_b3\_01} (red lines). Note that for the outflow rate, we show both the total outflow (solid lines) and the unbound component (dashed lines). Inflow/outflow rates are scaled by the peak injection rate given in Table \ref{tab:tab1}. The initial inflow rate of \texttt{e99\_b5\_01} is nearly twice that of \texttt{e99\_b3\_01}, reflecting the much more efficient circularization. The total mass outflow rates are similar throughout. However, the amount of gas that is unbound drops over time in \texttt{e99\_b3\_01} while \texttt{e99\_b5\_01} unbinds nearly 100\% of its outflow.}
    \label{fig:mdote99}
\end{figure}
%%%%%%%%%%
% End Figure
%%%%%%%%%

The dissipation of kinetic energy in the self-intersection shock causes the orbital binding energy of the shocked gas to decrease. This leads to the formation of a circularized accretion disk which has a lower eccentricity than the injected material. Secondary shocks in the forming disk \citep{Bonnerot2020,Bonnerot2021} and the stream disruption events \citep{Andalman2020} such as in Figure \ref{fig:e99beta5evol} cause additional dissipation but we do not explicitly track this dissipation rate. Instead, we use the eccentricity as a metric for the efficiency of binding energy dissipation. We track the eccentricity evolution of the disk material over the duration of each simulation by computing the mass weighted eccentricity. The eccentricity for each grid point is given by:
\begin{equation}
  e=\sqrt{1+2l^2 \epsilon},
\end{equation}
where $l=u_\phi$ is the specific angular momentum and $\epsilon = -(1+u_t)$ is the specific binding energy. We then compute the mass weighted eccentricity as a function of radius:
\begin{equation} \label{eq:meanecc}
  \langle e \rangle(r) = \dfrac{\int_0^{2\pi} \int_{\pi/3}^{2\pi/3} \sqrt{-g} \rho e \, d\theta \, d\phi}{\int_0^{2\pi} \int_{\pi/3}^{2\pi/3} \sqrt{-g} \rho\, d\theta \, d\phi}.
\end{equation}
We only integrate over a $\pm \pi/6$ wedge around the equatorial plane ($\theta=\pi/2$) which includes most of the forming disk. We quantify the disk thickness by estimating the density scale height over a $\pm\pi/4$ wedge around the equatorial plane:
\begin{equation}
    \dfrac{H}{R} = \dfrac{\int_0^{2\pi} \int_{\pi/4}^{3\pi/4} \rho \, \, \tan (| \pi/2 - \theta |)^2 \, d\theta d\phi}{\int_0^{2\pi} \int_{\pi/4}^{3\pi/4} \rho \,d\theta d\phi}.
\end{equation}

As shown in Figures \ref{fig:e99b5ecc} and \ref{fig:e99b3ecc}, the disk eccentricity decreases substantially over the course of a simulation. Similar to \citet{Andalman2020}, the injected stream constantly delivers gas with high eccentricity so the overall disk never reaches the approximate value for a highly circularized disk of $e\approx 0.3$ \citep{Bonnerot2016}. As expected based on the total energy dissipated at the circularization radius, \texttt{e99\_b3\_01} circularizes more slowly, decreasing below $e=0.8$ after $t=30,000\,t_g$. The lower eccentricity at radii lower than $r=16\,r_g$ is due to gas contained within the pericenter radius $R_p \approx 16\,r_g$ being relatively unmixed with gas that has yet to circularize. 

The accretion and outflow rates for each simulation are detailed in Figure \ref{fig:mdote99}. We compute the total inflow/outflow rate as:
\begin{equation} \label{eq:mdotin}
  \dot{M}_{\rm{in}}(r) = -\int_0^\pi \int_0^{2\pi} \sqrt{-g}\rho \,{\rm{min}}(u^r,0) d\phi d\theta.
\end{equation}
\begin{equation} \label{eq:mdotout}
  \dot{M}_{\rm{out}}(r) = \int_0^\pi \int_0^{2\pi} \sqrt{-g}\rho \,{\rm{max}}(u^r,0) d\phi d\theta.
\end{equation}
Note the extra $-1$ in the definition of the inflow rate since the integrand in this case is negative. In addition, we only consider fluid elements with positive Bernoulli number to contribute to the outflow since these gas parcels are expected to remain unbound as they travel to infinity. We define the Bernoulli number as:
\begin{equation} \label{eq:Be}
  {\rm{Be}} = -\dfrac{T^t_{\ \, t} + R^t_{\ \, t} + \rho u^t}{\rho u^t}.
\end{equation}
The density of the outflow is substantial and it remains optically thick for the duration simulated in this work. We describe the photosphere in a later section.

The accretion rate of mass crossing the BH horizon is several times the Eddington rate in both \texttt{e99\_b5\_01} and \texttt{e99\_b3\_01} (Figure \ref{fig:mdote99}). The accretion rate of \texttt{e99\_05\_01} is nearly twice that of \texttt{e99\_03\_01} initially, reflecting that the dissipation of orbital energy is more rapid for this closer disruption. The accretion rate grows as the mean eccentricity of the disk decreases and appears to saturate after $t=30,000\,t_g$ for \texttt{e99\_b5\_01}, which is approximately when the eccentricity in the inner disk reaches its lowest value of $e\approx 0.6$ (Figure \ref{fig:e99b5ecc}). In \texttt{e99\_b3\_01}, this saturation appears to occur slightly later at $t=50,000 t_g$. This increase in accretion rate also appears to correlate to the system approaching an inflow/outflow equilibrium. Early in the disk formation, the total mass inflow rate as a function of radius for $R_{\rm{min}}<r<R_p$ is constant, but smaller than the net inflow rate at radii $R_p<r<R_{\rm{inj}}$. At the point that the accretion rate saturates ($t>30,000\, t_g$ for \texttt{e99\_b5\_01} and $t>50,000\, t_g$ for \texttt{e99\_b3\_01}), the total inflow rate at all radii is nearly in equilibrium (i.e. nearly constant) for $R_{\rm{min}}<r<R_{\rm{inj}}$.

%%%%%%%%
%Begin Figures
%%%%%%%%
\begin{figure}
    \centering{}
	\includegraphics[width=\columnwidth]{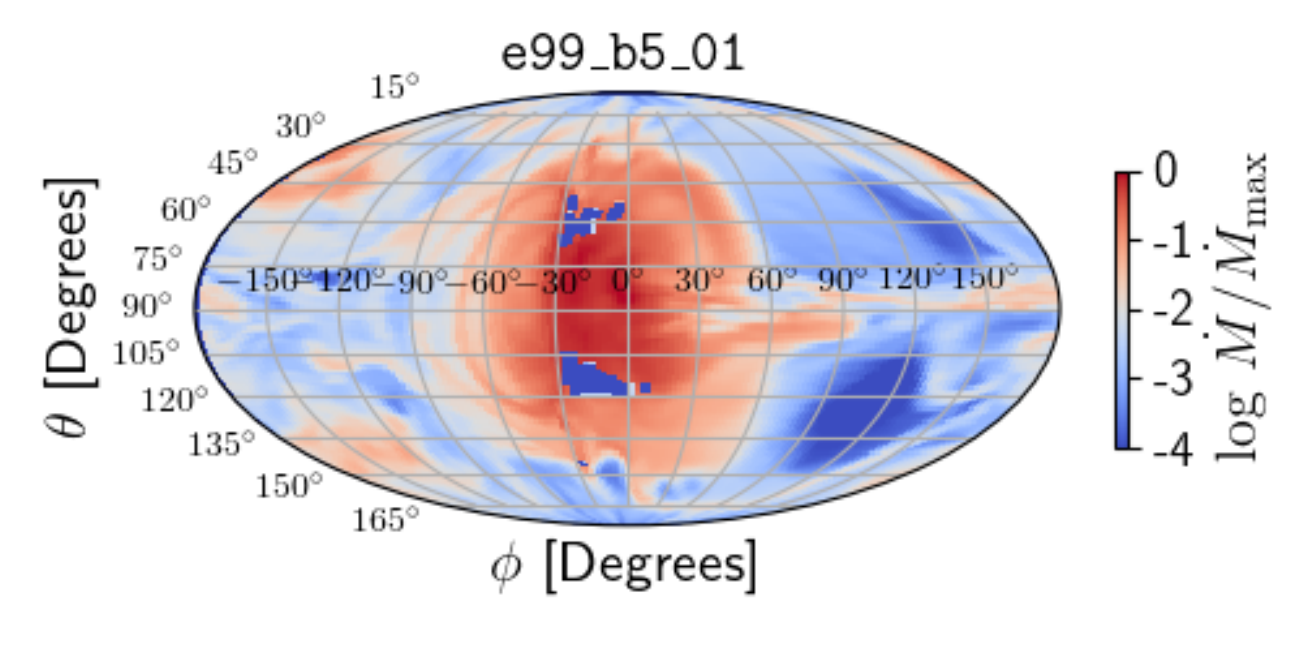}\\
	\includegraphics[width=\columnwidth]{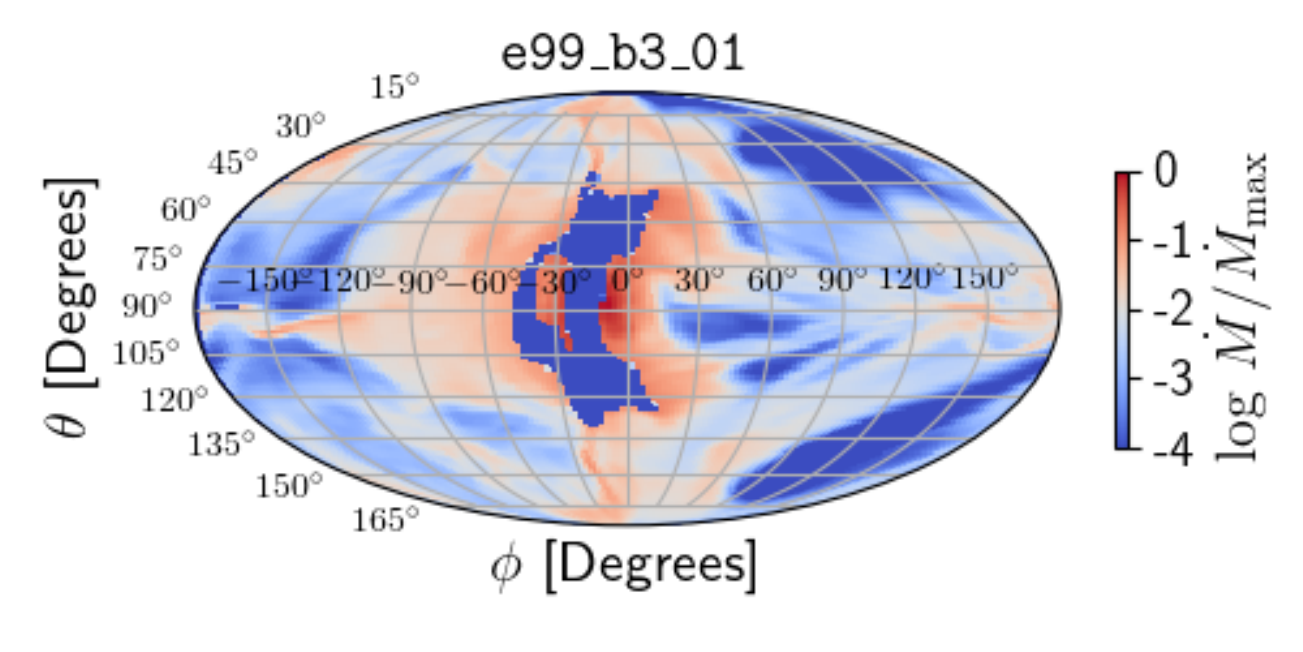}
    \caption{Here we show the mass outflow of unbound gas ($Be>0$) for \texttt{e99\_b5\_01} (top panel) and \texttt{e99\_b3\_01} (bottom panel) at each point on a spherical surface at $r=1000\, r_g$ for a snapshot of the simulation at $t=30,000\,t_g$. We normalize the outflow rate at each point by the maximum outflow rate on the surface.}
    \label{fig:mdout}
\end{figure}

\begin{figure}
    \centering{}
	\includegraphics[width=\columnwidth]{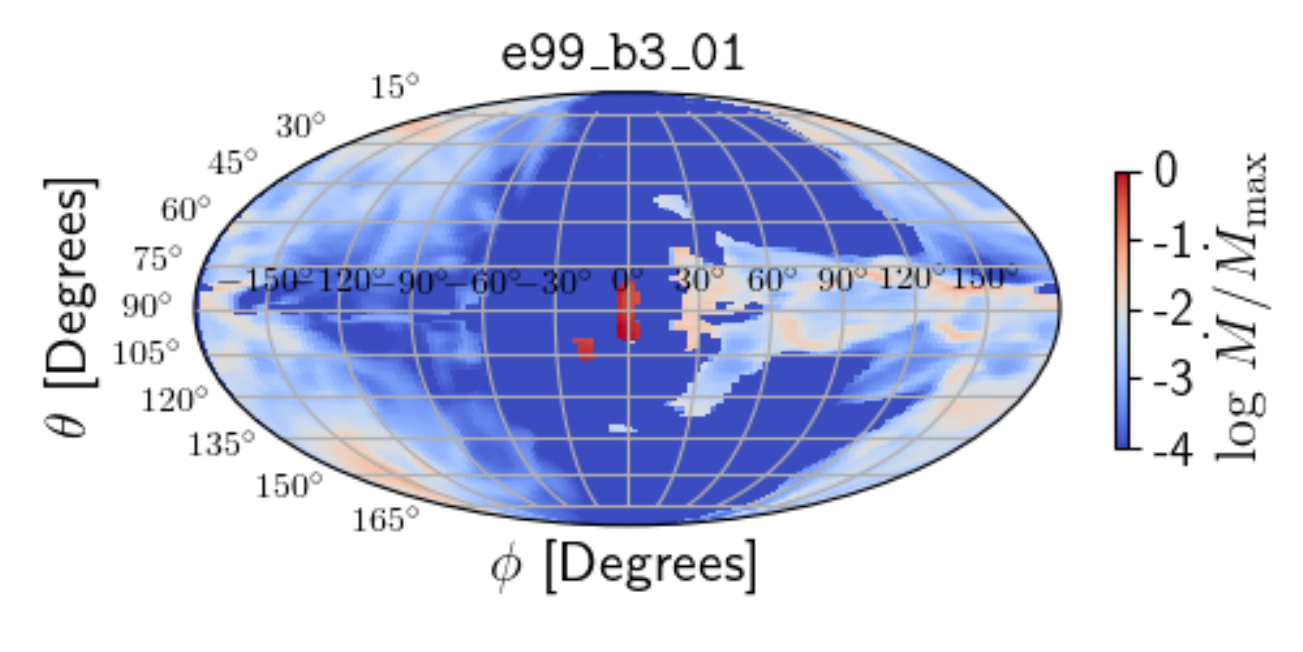}
    \caption{The same as Figure \ref{fig:mdout} but for \texttt{e99\_b3\_01} at $t=60,000\,t_g$.}
    \label{fig:e99b3outflowend}
\end{figure}
%%%%%%%%
%End Figures
%%%%%%%%

\subsection{Outflows} \label{sec:e99outflows}

We show the outflow crossing the shell at $r=1000\,r_g$ in the bottom panel of Figure \ref{fig:mdote99}. The self intersection of the stream leads to a significant fraction of the shocked gas becoming unbound, i.e. $\dot{M}_{\rm{out}}$ peaks at $\approx60\dot{M}_{\rm{Edd}}$ compared to the peak injection rate of $133\dot{M}_{\rm{Edd}}$. As we demonstrate in \S\ref{sec:resultse97}, radiation is able to diffuse through the surrounding gas and drives even more gas to become unbound. The typical velocity of the outflow is $v\approx 0.1c$.

Periodic behavior is exhibited in the outflow rate of \texttt{e99\_b5\_01} on top of the overall long term trend. The period is of the order $P\approx 3300\,t_g$ and the amplitude of the variation is nearly $10\dot{M}_{\rm{Edd}}$. As was noted in \citet{Sadowski2016a}, this periodic behavior is due to the large angular momentum transfer in the self-intersection region to the part of the stream that is making its first return to pericenter. This sets up a feedback loop. The gas that has already passed pericenter and precessed, not only causes a shock at the self-intersection point but also deposits angular momentum. This pushes the incoming gas out to larger orbits, which leads to weaker precession and then subsequent self intersection at larger radii. However, self intersections at larger radii transfer momentum less efficiently, so the incoming stream is then able to return to its original orbit and undergo stronger relativistic orbital precession, thus resetting the feedback loop. The period of the feedback loop is determined by the radius at which the collision occurs. For \texttt{e99\_b5\_01}, the Keplerian radius associated with the feedback period is $\approx 65\,r_g$, which falls within the region we identify with self-intersection (Figure \ref{fig:e99beta5hm}). For \texttt{e99\_b3\_01}, the period is  $P\approx 8400\,t_g$ which corresponds to a Keplerian radius of $\approx121\,r_g$.

We note that our simulations appear to predict a rather large outflow at the peak outflow rate (nearly 45\% of the injected mass) for a $10^6\,M_\odot$ mass BH. This is owing to the larger impact parameter ($\beta=3, ~5$) in our simulations, which means that more kinetic energy is available for dissipation in the self intersection. \citet{Lu2020} show that close disruptions launch a more energetic, higher velocity outflow and that the fraction of gas in the outflow that becomes unbound is sensitive to the impact parameter. They predict that the critical BH mass above which more than 20\% of the inflowing gas becomes unbound for a $\beta=5$ disruption is $M_{\rm{cr}}\approx3\times10^5\,M_\odot$ while for a $\beta=3$ disruption $M_{\rm{cr}}\approx7\times10^5\,M_\odot$. While our simulations exceed this estimated critical mass, we caution that \citet{Lu2020} provide estimates based on streams with the binding energy for an $e=1$ disruption. Our streams are more bound owing to the choice of $e=0.99$, so the corresponding critical mass for this work is likely slightly higher than the values above. Nevertheless, the fact that both simulations initially unbind far more than 20\% of the shocked gas suggests that the corresponding critical mass for our choice of parameters is lower than the BH mass of $10^6\,M_\odot$ that we employ. The precise fraction of gas that becomes unbound is not provided by \citet{Lu2020} for the parameters in the present work; however, \texttt{e99\_b5\_01} and \texttt{e99\_b3\_01} appear to eject a similar amount of mass at the peak outflow rate (Figure \ref{fig:mdote99}) despite the difference in impact parameter, $\beta$. 

We note that it is possible that the use of the Bernoulli number to track unbound gas could in principle over estimate the mass of unbound gas in the outflow since the radiation component could simply escape once the gas becomes optically thin and not get deposited in kinetic energy. However, we find that the specific binding energy alone is net positive in the outflow where regions with positive Bernoulli have been identified, so we find that this result is consistent regardless of whether or not radiation escapes.

We perform a Mollweide projection of the unbound outflow through a spherical shell at radius $r=1000r_g$ to display the angular distribution of the outflow for a snapshot of the simulation at $t=30,000\,t_g$ (Figure \ref{fig:mdout}). In the figure, the stream injection point is located at $\theta=90^\circ$, $\phi=0^\circ$, while pericenter occurs at $\phi\approx 180^\circ$. The poles are situated at $\theta = 0^\circ$ and $180^\circ$. We find that the majority of the unbound outflow during the peak outflow rate is directed radially away from the self-intersection point, roughly back towards the stream injection point, but subtending a significant solid angle around this direction. There is also a significant amount of gas flowing near the poles and near pericenter but it contributes less than 10\% of the total outflowing mass at the peak of the outflow. 

While the top panel of Figure \ref{fig:mdout} is representative of the unbound outflow in \texttt{e99\_b5\_01} throughout its evolution, the outflow centered on $\theta\approx 90^\circ, \, \phi\approx0^\circ$ in Figure \ref{fig:mdout} is largely bound by the end of \texttt{e99\_b3\_01} as shown in Figure \ref{fig:e99b3outflowend}. This change is also apparent in Figure \ref{fig:mdote99}. While nearly half of the injected gas becomes unbound throughout the entire evolution of \texttt{e99\_b5\_01}, in the case of \texttt{e99\_b3\_01} the outflow rate of unbound gas drops to nearly 5-10\% of the mass injection rate by the end of the simulation (Figure \ref{fig:mdote99}). This change in behaviour is due to the stream deflection described above. Due to the lower $\beta$, the change in collision radius during periods of stream deflection lead to a large enough decrease in dissipated kinetic energy as to substantially decrease the fraction of mass that becomes unbound.

\subsection{Radiation Properties} \label{sec:photosphere}

%%%%%%%%%%%%%%%%%%%%%%%%%%%%% Figures
\begin{figure}
    \centering{}
	\includegraphics[width=\columnwidth]{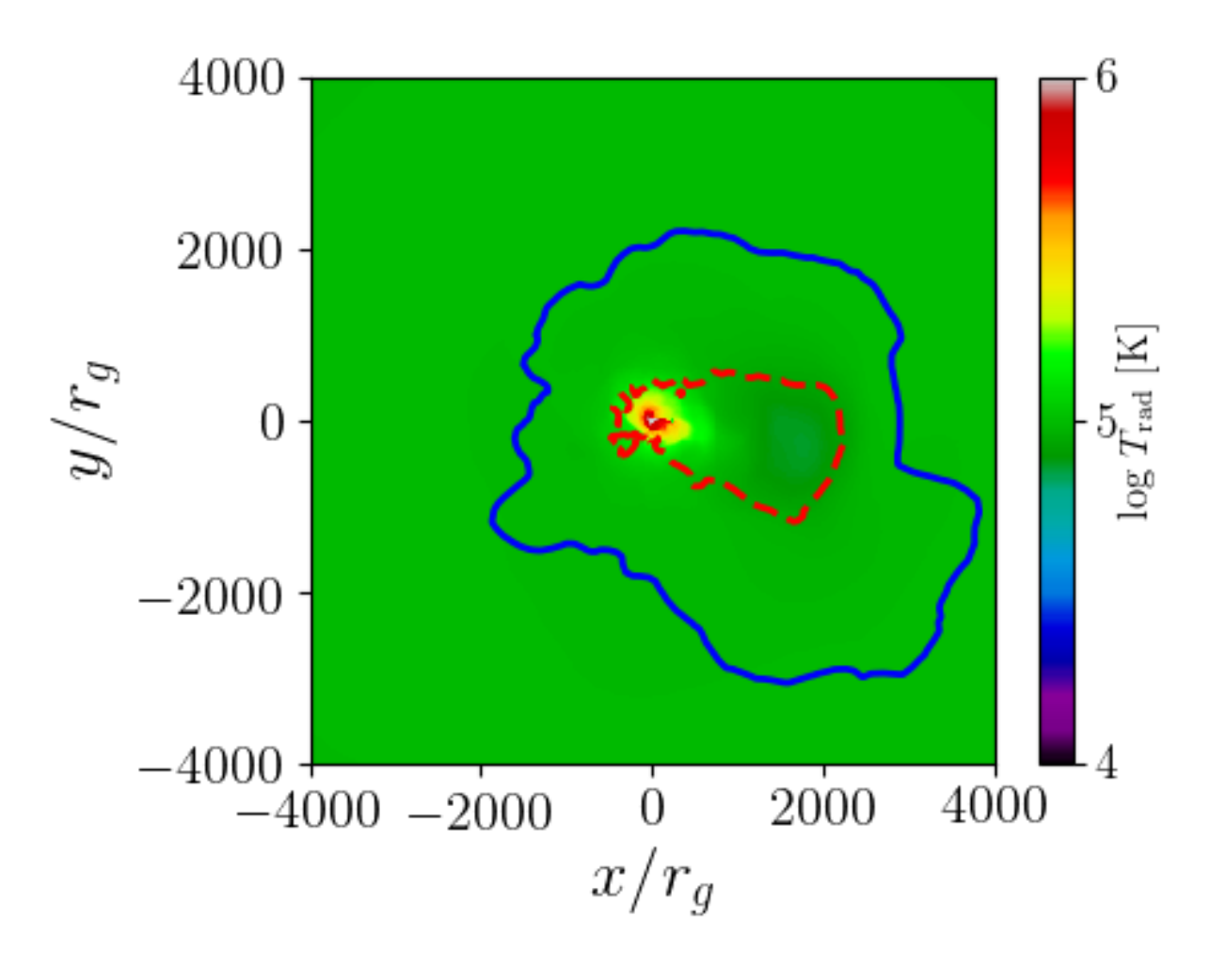}\\
	\includegraphics[width=\columnwidth]{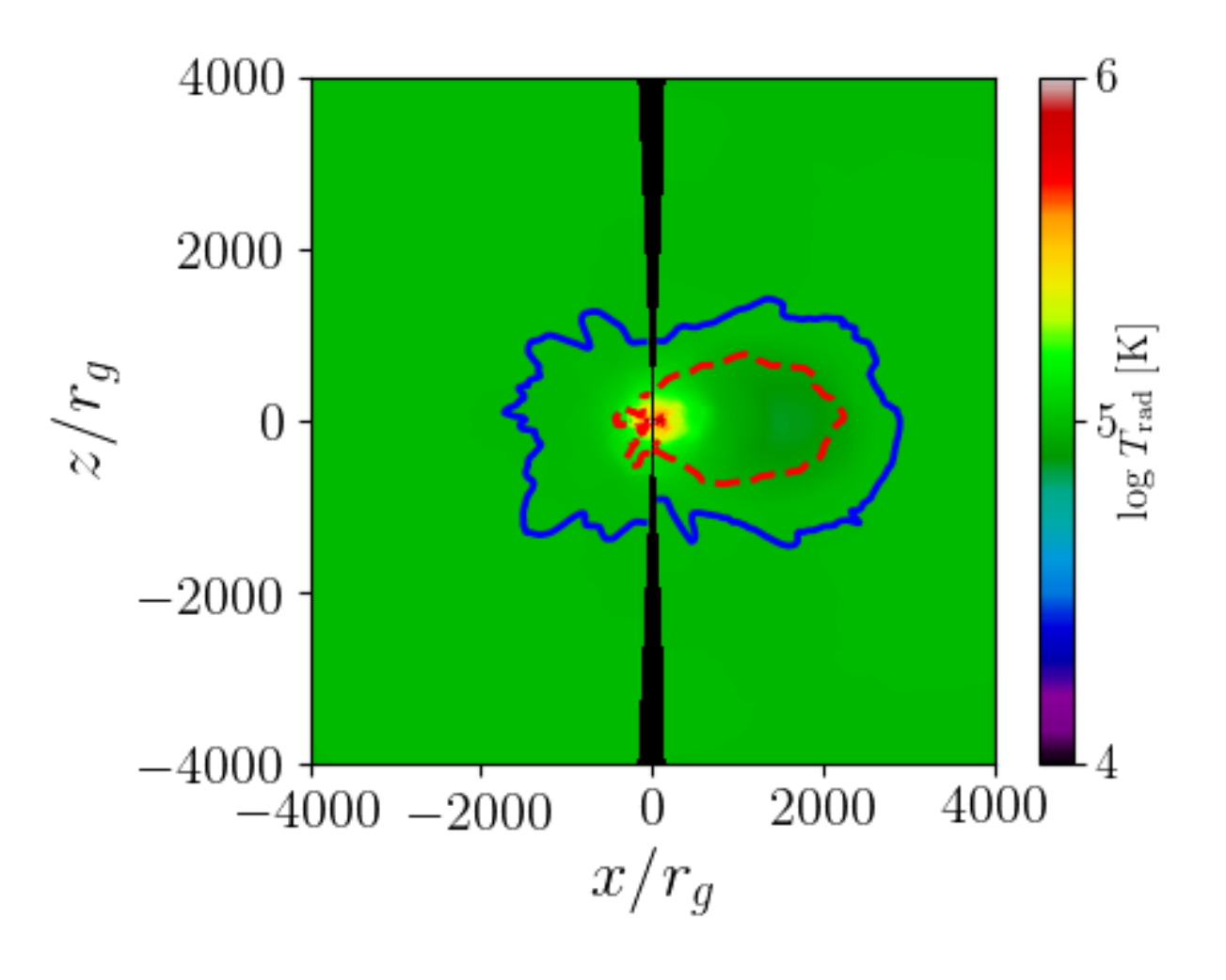}
    \caption{Here we depict the radiation temperature and photosphere of \texttt{e99\_b5\_01} at $t=60,000\, t_g$. We show cross sections of the radiation temperature (colors), photosphere (blue line), and photon trapping surface (red dashed line) for the equatorial plane (top panel) and an aziumuthal slice that intersects the injection point and pericenter (bottom). The stream is injected at $x=200r_g$, $y=z=0$.}
    \label{fig:tauphot}
\end{figure}

\begin{figure}
    \centering{}
	\includegraphics[width=\columnwidth]{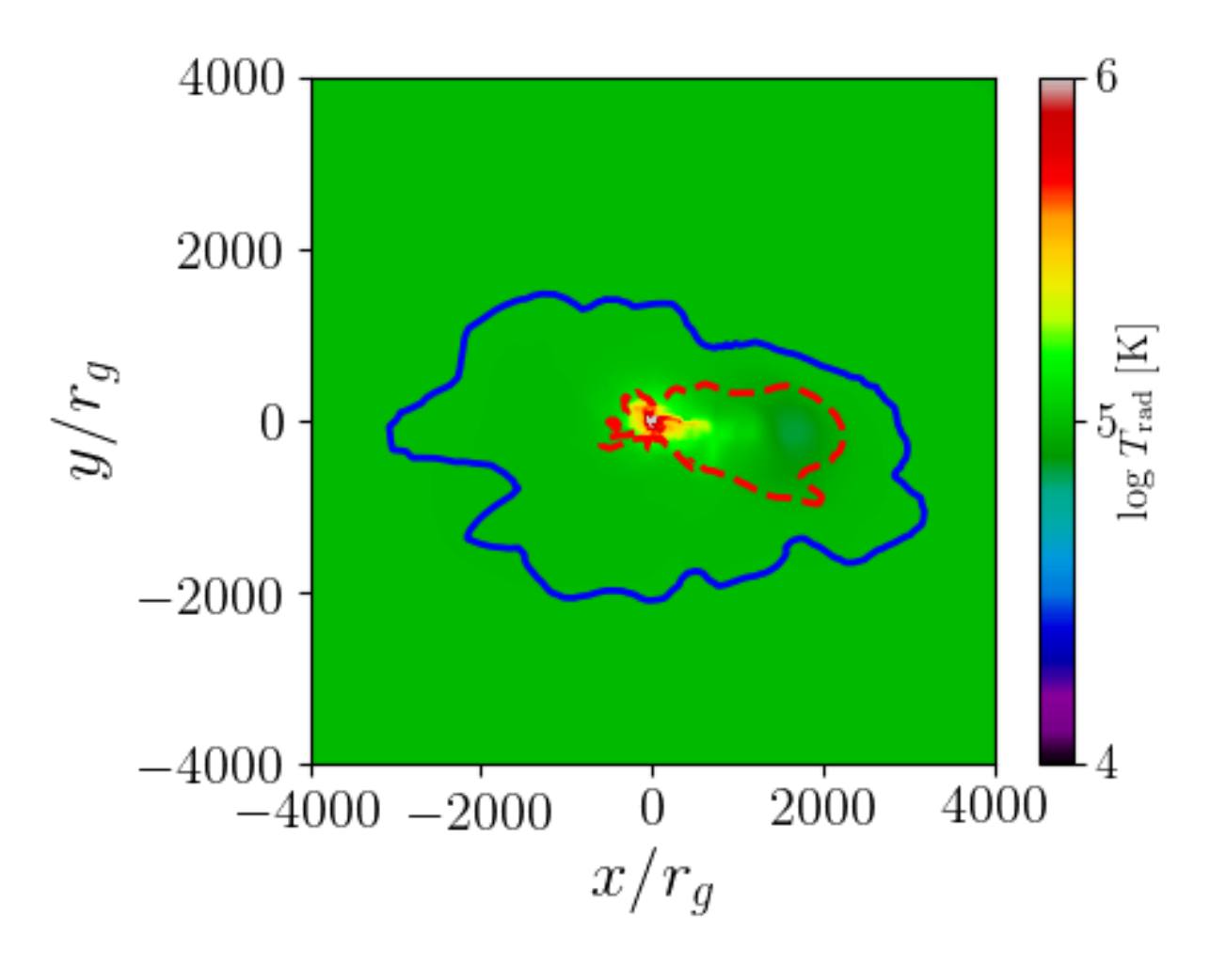}\\
	\includegraphics[width=\columnwidth]{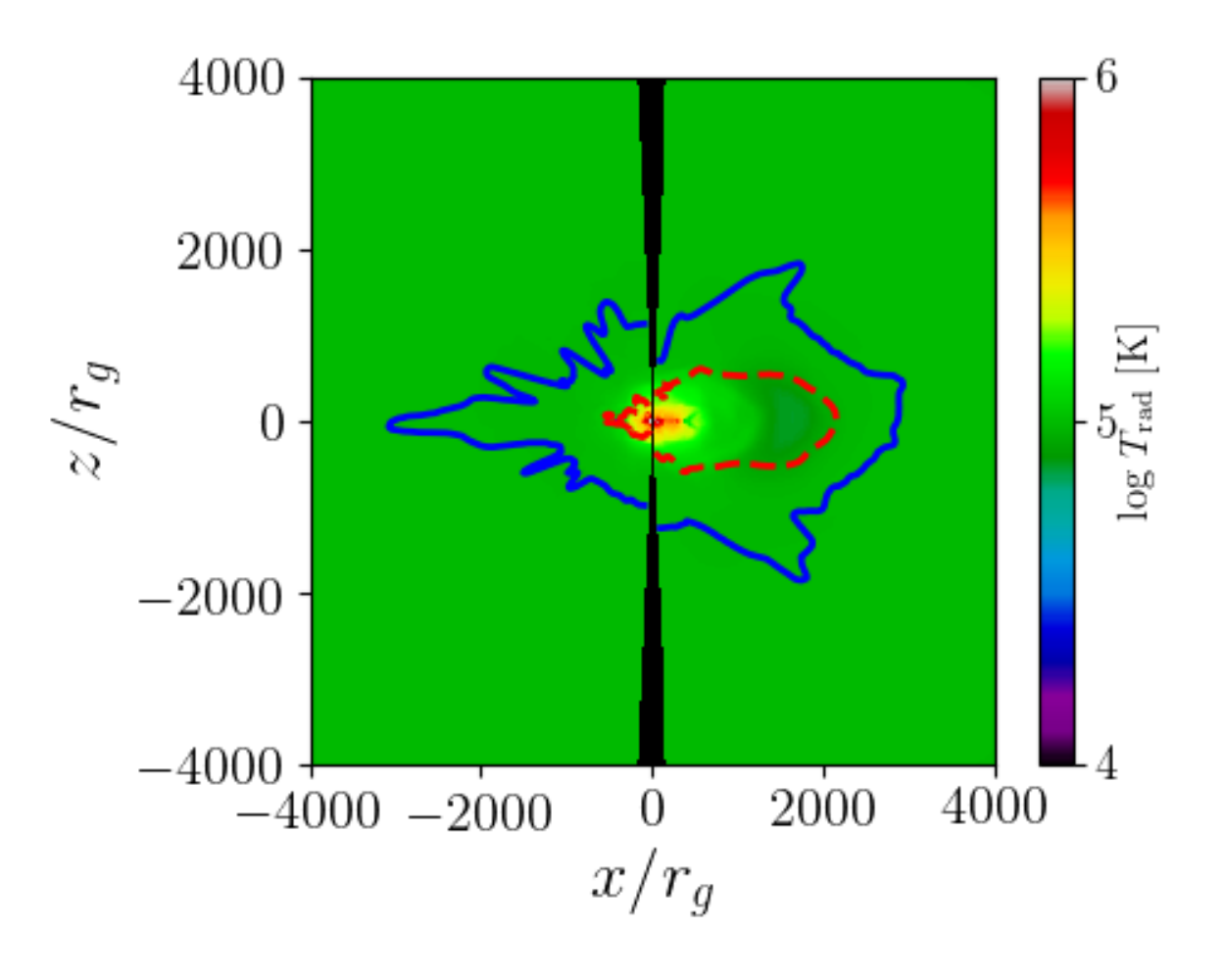}
    \caption{The same as Figure \ref{fig:tauphot} but for \texttt{e99\_b3\_01} at $t=60,000\,t_g$. The stream is injected at $x=400r_g$, $y=z=0$.}
    \label{fig:tauphote99b3}
\end{figure}

\begin{figure}
    \centering{}
	\includegraphics[width=\columnwidth]{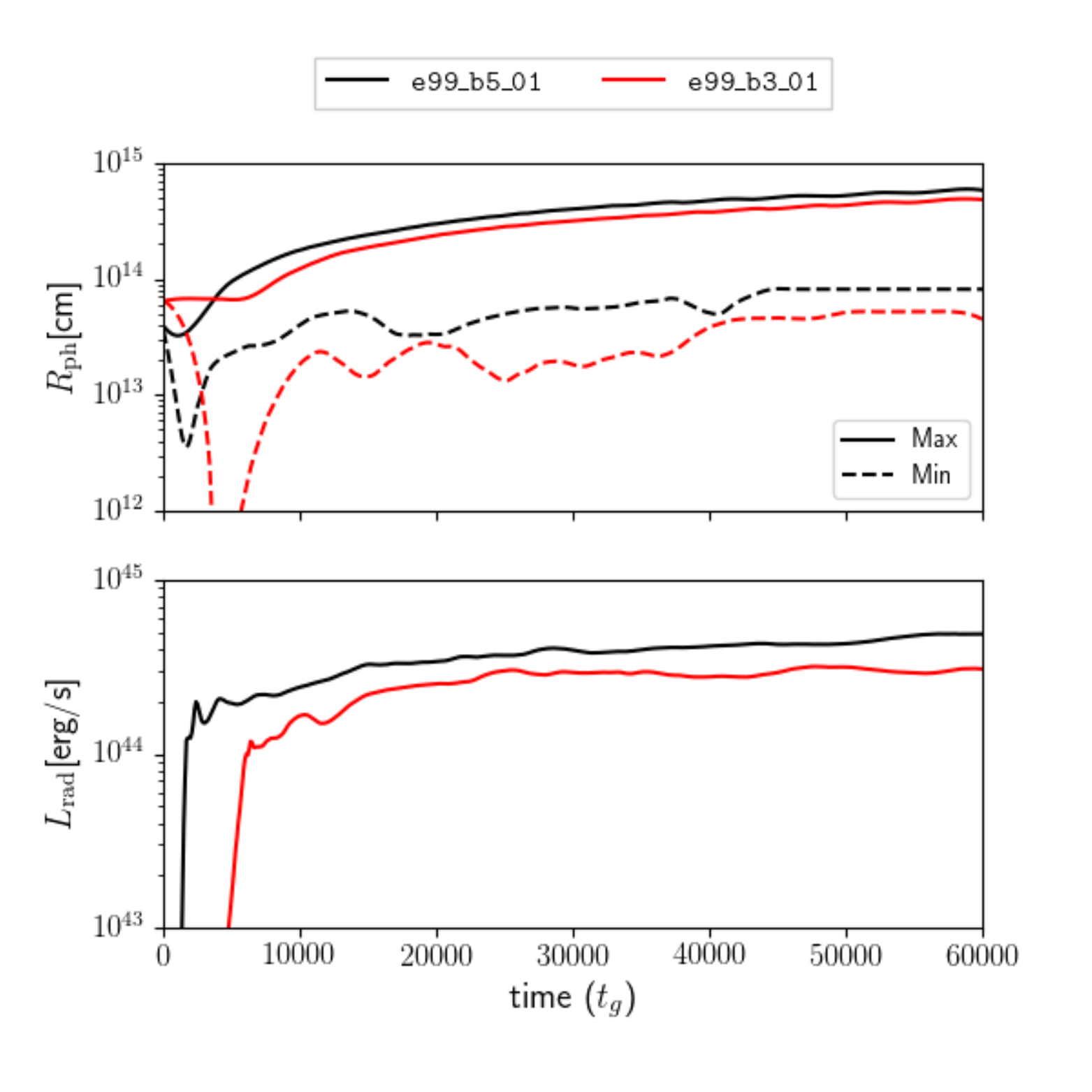}
    \caption{Here we show the photosphere radius (top) and bolometric luminosity (bottom), both in physical units, for \texttt{e99\_b5\_01} and \texttt{e99\_b3\_01}. In the top panel, we show the minimum (dashed line) and maximum (solid line) photosphere radius over time. The minimum photosphere radius (top panel) shows occasional dips during epochs where gas at the poles is infalling and the photosphere radius consequently decreases.}
    \label{fig:lrad}
\end{figure}

\begin{figure}
    \centering{}
	\includegraphics[width=\columnwidth]{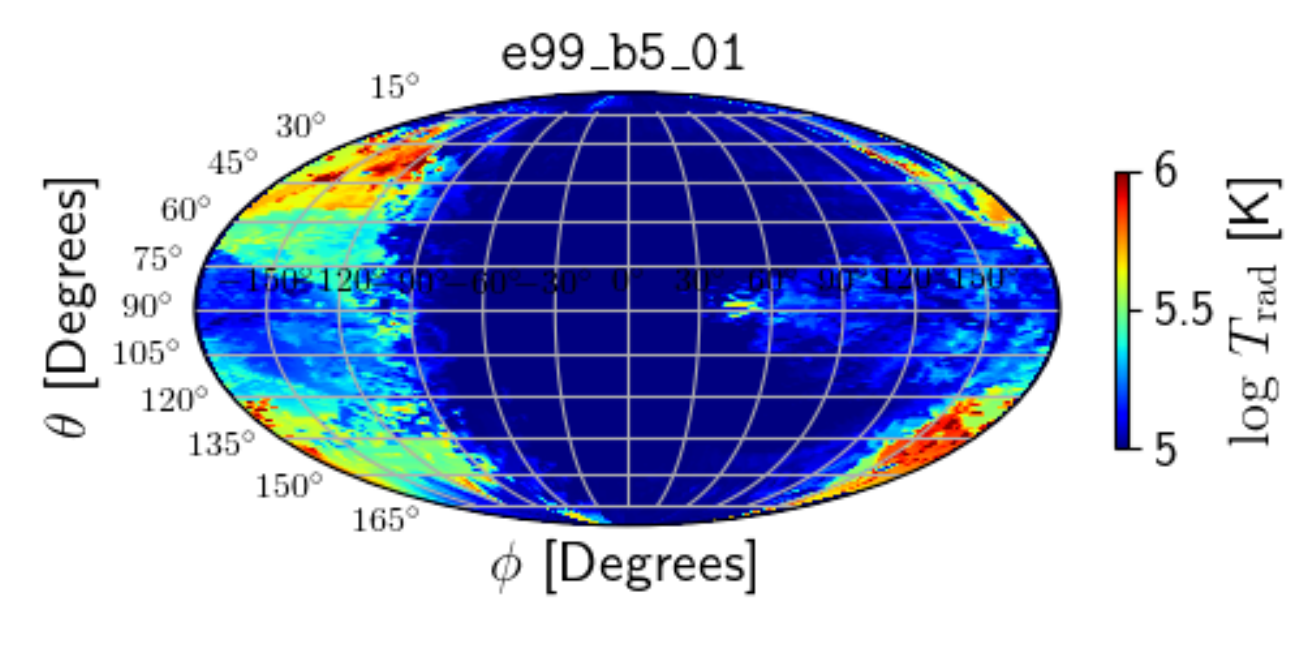}\\
	\includegraphics[width=\columnwidth]{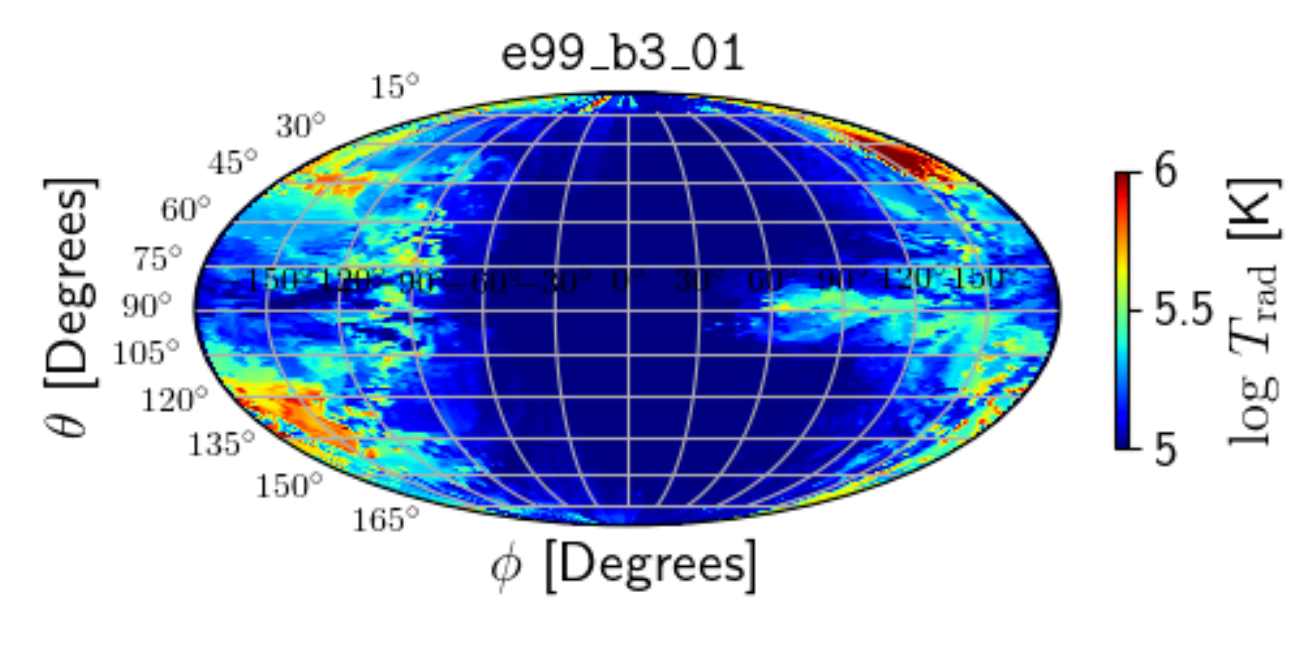}
    \caption{Here we show the projection of the radiation temperature at the thermalization surface ($R_{\rm{th}}$) for \texttt{e99\_b5\_01} (top panel) and \texttt{e99\_b3\_01} (bottom panel) along each line of sight at $t=60,000\,t_g$. The figure is discussed in the text.}
    \label{fig:Tradprojection}
\end{figure}
%%%%%%%%%%%%%%%%%%%%%%%%%%%%% End Figures

As we discuss in \S\ref{sec:resultse97}, radiation plays an important role in the gas dynamics. The typical picture in super-Eddington accretion flows is that radiation is trapped and advected with the gas within the optically thick accretion flow, but it can diffuse more effectively in the outflows. We estimate the diffusion timescale in the accretion disk and outflow to confirm this picture. 

For the accretion disk in \texttt{e99\_b5\_01} and \texttt{e99\_b3\_01}, we estimate the time scales for diffusion and advection within the disk (i.e. regions within $H/R=0.4$ for radii $r<100\,r_g$) as $t_{\rm{dif,disk}}\approx 3\tau_z (H/R)R$ and $t_{\rm{adv}}=R/v_{\rm{in}}$. More succinctly, $t_{\rm{dif,disk}}/t_{\rm{adv}}\approx 3\tau_z (H/R) v_{\rm{in}}$. Here $(H/R)$ is the density scale height of the disk (which is approximately $0.4$, shown in Figures \ref{fig:e99b5ecc} and \ref{fig:e99b3ecc}), $\tau_z$ is the vertically integrated optical depth through the disk, and $v_{\rm{in}}$ is the inflow velocity. The opacity is estimated using the Thomson scattering opacity, which is $\kappa_{\rm{es}}=0.34$ cm$^2$ g$^{-1}$ for Solar metallicity. We compute the vertically integrated optical depth as:
\begin{equation} \label{eq:verttau}
    \tau_z = \int_{0}^{H} \rho \kappa_{\rm{es}}\, dz.
\end{equation}
For all times after the disk has begun to form (i.e. after the initial peak in the accretion rate in Figure \ref{fig:mdote99}), the general description of the diffusion and advection times in the following calculation holds. 

For radii within $r<20 r_g$ the inflow velocity, $v_{\rm{in}}$, increases towards the BH horizon with a minimum value of $2\times10^{-2}c$ and a maximum of $0.6c$. For $r>20\, r_g$ the inflow velocity is nearly constant with a value of $2\times10^{-2}c$. The vertically integrated optical depth for both \texttt{e99\_b5\_01} and \texttt{e99\_b3\_01} ranges between $200<\tau<500$ in the forming disk. As a consequence, we find that $t_{\rm{diff}}/t_{\rm{adv}} \approx 4.8-12$ for radii $r>20r_g$ while for $r<20r_g$ the ratio increases rapidly. While the ratio is not significantly larger than unity, it suggests that within the forming disk, since the inflow velocity is quite large, advection is the primary radiative transport mechanism. 

Outside of the inner accretion disk, the dynamical time is $t_{\rm{dyn}}=v/R$ where $v$ is the gas velocity and the diffusion time is estimated using the radially integrated optical depth:
\begin{equation} \label{eq:tauradial}
    \tau_{\rm{es}}(R) \equiv \int_{R}^{R_{\rm max}} \rho \kappa_{\rm{es}}\, dr,
\end{equation}
such that $t_{\rm{dif,outflow}}=\tau_{\rm{es}}(R) R$. The structure of the outflow's trapping surface, where the diffusion and dynamical time are equal, is quite asymmetrical (red dashed line in Figures \ref{fig:tauphot} and \ref{fig:tauphote99b3}). Note that for regions interior to the trapping surface contour $t_{\rm{diff}}/t_{\rm{adv}} > 1$ while the opposite is true outside of it. In general, gas near pericenter exhibits a trapping surface that is close to the radial boundary of the forming disk ($100-200 r_g$). Outside of this surface, radiation can decouple from the gas. On the opposite side of the BH the trapping surface is almost as distant as the photosphere. The fact that gas is able to diffuse near pericenter perhaps explains the weak outflows of gas near pericenter ($\phi\approx \pm \pi$) in Figure \ref{fig:mdout}.

We begin our discussion of the emitted radiation by describing the photosphere. At each $(\theta,\phi)$, we integrate radially inward from $R_{\rm{max}}$ to find the photosphere radius $R_{\rm ph}$ defined by: 
\begin{equation}
    \tau_{\rm{es}}(R_{\rm ph}) \equiv \int_{R_{\rm{ph}}}^{R_{\rm max}} \rho \kappa_{\rm{es}}\, dr = \frac{2}{3},
\end{equation}
where $R_{\rm ph}$ is the radius at which the optical depth is equal to $2/3$. As we show in Figures \ref{fig:tauphot} and \ref{fig:tauphote99b3}, the photosphere of \texttt{e99\_b5\_01} and \texttt{e99\_b3\_01} is highly asymmetric and irregularly shaped. The radiation temperature at the photosphere maintains a nearly constant value of $10^5$ K over the simulation. In general, the photosphere radius is closest to the accretion disk near pericenter and near the poles while on the other side of the BH, the side where the self-intersection occurs, the photosphere radius is much larger.

We obtain the bolometeric luminosity directly from the radiation stress energy tensor by integrating over the photosphere. The radiation luminosity is taken as all outgoing rays of radiative flux at the electron scattering photosphere ($R_{\tau=2/3}$):
\begin{equation}
    L_{\rm{bol}} = -\int_0^{2\pi} \int_0^\pi \sqrt{-g}R^r_t \, d\theta d\phi.
\end{equation}
These rays are assumed to reach a distant observer. We show the minimum/maximum photosphere radius as well as the radiant luminosity in Figure \ref{fig:lrad}. The bolometric luminosity is mildly super-Eddington in both \texttt{e99\_b5\_01} and \texttt{e99\_b3\_01}. We note that \texttt{e99\_b3\_01} exhibits a slightly lower luminosity over its evolution owing to the less energetic self intersection which largely characterizes the energetics of the event.

The radiation temperature at the photosphere of $\lesssim 10^5$\,K is significantly hotter than that observed in optically identified TDEs, where typical temperatures are on the order of $\sim (1-{\rm few})\times 10^4$ K. However, the duration of time that we simulate, which is only 3.5 days, may be more analogous to the beginning of the flare during the rise to peak. For instance, ASASSN-19bt exhibited bright UV emission with a temperature peak of $T\approx 10^{4.6}$ K which then decayed to $T\approx 10^{4.3}$ K over several days \citep{Holoien2019}. This may indicate that some TDEs in fact start out with hotter emission and quickly cool as the photosphere expands. 

The geometry of the photosphere is particularly interesting. We find that the minimum photosphere radius (which occurs close to pericenter) is only $R_{\rm{ph}}\approx 3-6\times10^{13}$ cm above and below the disk at late times (see Figure \ref{fig:lrad}). This may be an ideal geometry for viewing angle dependent X-ray emission. The radiation temperature in Figure \ref{fig:tauphote99b3} is only approximate without detailed radiative transfer, and for such small photosphere radii in the pericenter direction it may be possible for X-rays to reach the photosphere before being absorbed, thus emerging as visible radiation. Meanwhile, for an observer viewing the photosphere from the equatorial plane at the point where the photosphere radius is largest, the X-rays are expected to be completely absorbed.

The radiation temperature in the inner accretion flow reaches $T_{\rm{rad}}\approx 10^6$ K. In regions where there is not much absorption, hot emission may diffuse and reach the scattering surface. As we do not carry out detailed ray tracing to determine the frequency dependent spectrum of the accretion flow, we estimate the emerging photon energy by accounting for the effects of bound-free absorption along radial trajectories. We did not directly include the effects of bound-free absorption during the evolution of the simulation, so we perform a post processing of the simulation data taking this additional source of opacity into account. We adopt the gray approximation of the absorption due to metals ($\kappa_{\rm{bf}}$) in the atmosphere via the model of \citet{Sutherland1993} and assume a Solar metal abundance for the gas. To test the possibility of X-ray emission, we find the thermalization radius ($R_{\rm{th}}$) by computing $\tau_{\rm{eff}}= \int_{R_{\rm{th}}}^{\infty} \rho \sqrt{\kappa_{\rm{abs}}(\kappa_{\rm{es}} + \kappa_{\rm{abs}})}\,dr $, where $\kappa_{\rm{abs}}=\kappa_{\rm{bf}}+\kappa_{\rm{ff}}$. In general, the thermalization radius is smaller than the photosphere radius and near pericenter it comes within $< 100\,r_g$ of the inner accretion flow along some lines of sight. We perform a Mollweide projection of the radiation temperature ($T_{\rm{rad}}$) at the thermalization radius as a function of viewing angle for \texttt{e99\_b5\_01} and \texttt{e99\_b3\_01} at $t=60,000\,t_g$ (Figure \ref{fig:Tradprojection}). In regions where the most gas blocks the line of sight and the photosphere radius is near its maximum ($\phi\sim 0^\circ$ in Figure \ref{fig:Tradprojection}), the radiation temperature is $T_{\rm{rad}}\approx 10^5$ K. Near pericenter ($\phi\sim \pm 180^\circ$ in Figure \ref{fig:Tradprojection}), and especially above/below the equatorial plane, the radiation temperature is much hotter and reaches a typical temperature of $T_{\rm{rad}} \approx 3-5\times 10^5$ K and some regions reach $10^6$ K. This treatment is only approximate, but suggests that close ($\beta\geq3$) TDEs around lower mass BHs may be sources of soft X-rays near the peak emission.

The radiative efficiency for each simulation is computed as:
\begin{equation}
 \eta = \eta_{\rm{NT}} \left(\dfrac{L_{\rm{bol}}}{L_{\rm{Edd}}}\right) \left(\dfrac{\dot{M}_{\rm{in}}}{\dot{M}_{\rm{Edd}}}\right)^{-1},
\end{equation}
where $\eta_{\rm{NT}}$ is defined in Equation \ref{eq:etaNT}. We use the average luminosity and accretion rate for the final $5,000\, t_g$ for each simulation. For \texttt{e99\_b5\_01}, the radiative luminosity is of the order $L_{\rm{bol}}\approx 5\, L_{\rm{Edd}}$ and the mean accretion rate is $\dot{M}_{\rm{in}}\approx 20\,\dot{M}_{\rm{Edd}}$. The estimated efficiency is then $\eta \approx 0.15\eta_{\rm{NT}}\approx 0.014$. For \texttt{e99\_b3\_01}, we find  $\eta \approx 0.15\eta_{\rm{NT}}\approx 0.009$ using a similar approach.

An ongoing curiosity of optically identified TDEs is that they appear to be either extremely radiatively inefficient, or they only accrete a small amount (some TDEs suggest only 1\% at minimum) of the stellar mass that is bound to the BH \citep{Holoien2014,Holoien2019,Holoien2020}. An interesting example is ASASSN-14ae, for which \citet{Holoien2014} estimate the mass needed to power the observed line emission is at least an order of magnitude higher than the minimum mass accretion to power the continuum emission, suggesting a lower radiative efficiency. Our simulation suggests that around 10-20\% of the inflowing material actually manages to accrete via an accretion flow while the disk is forming. If the prompt emission from optical TDEs is thermal emission from the self intersection outflow, the radiative efficiency is indeed expected to be low.

We note that the omission of the magnetic fields in \texttt{e99\_b5\_01} and \texttt{e99\_b3\_01} may impact the above result as the turbulence sourced by the MRI may lead to higher accretion rates. However, as we discuss in \S\ref{sec:resultse97}, our runs which included the magnetic field replicate the results obtained by \citet{Sadowski2016a}, who showed that hydrodynamical viscosity dominates the gas dynamics.

%%%%
% End of Section 4
%%%%

\section{Less Eccentric Models} \label{sec:resultse97}

In the previous section we discussed our primary simulations, \texttt{e99\_b5\_01} and \texttt{e99\_b3\_01}, which correspond to tidal disruptions of stars on highly eccentric orbits with $e=0.99$. Here we discuss briefly the evolution of less eccentric models with $e=0.97$ and $\beta=5$. These simulations were performed to compare the method of injection with previous work done using more bound stars and to illustrate the effects of radiation in comparison to pure hydrodynamics. In \texttt{e97\_b5\_01}, we include the magnetic field to examine if the magnetic field becomes dynamically important. We also set the spin of the BH to $a_*=0.9$ to confirm that a jet is not produced during the disk formation if a weak field is present in the TDE stream. We compare it with \texttt{e97\_b5\_02} to illustrate the impact of radiation for extremely optically thick TDE disk simulations when radiation is included versus pure hydrodynamics. In \texttt{e97\_b5\_03} and \texttt{e97\_b5\_04}, the disruption properties are the same as in \texttt{e97\_b5\_01} and \texttt{e97\_b5\_02} but we inject only $0.04\,M_\odot$. These simulations illustrate the impact of radiation in less dense atmospheres on the outflow and accretion rate. We also compare the disk properties with previous simulations of TDE disks.

We note that the magnetic field is not relevant for comparison with models described in \S\ref{sec:resultse99} nor other simulations in this section, as will be discussed later in this section.

\subsection{Dynamics}

As described in \S\ref{sec:nummethods}, the TDE stream is injected for a finite amount of time. In the simulations of $e=0.97$ TDEs, the tail end of the stream is injected at $t=\Delta t_{\rm{inj}}=14,467 \,t_g$, but we evolve the simulation beyond this point. The evolution during the stream injection phase is similar to the higher eccentricity models discussed in \S\ref{sec:resultse99}. We note that due to the lower eccentricity, the stream thickness in the orbital plane is slightly larger than the simulations discussed in \S\ref{sec:resultse99}. This leads to slightly more expansion of the gas as it passes through pericenter.

\begin{figure}
    \centering{}
	\includegraphics[width=\columnwidth]{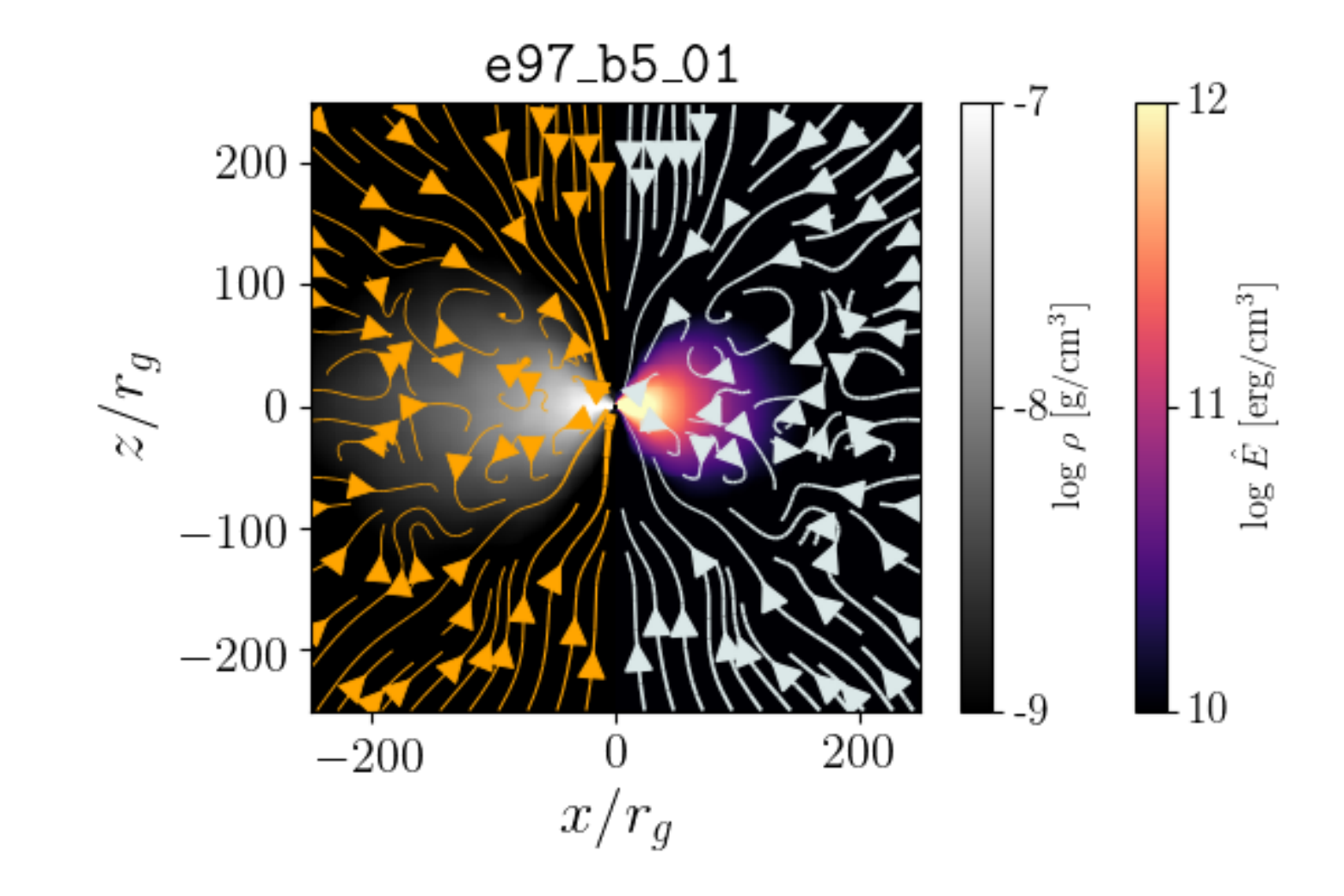}\\
	\includegraphics[width=\columnwidth]{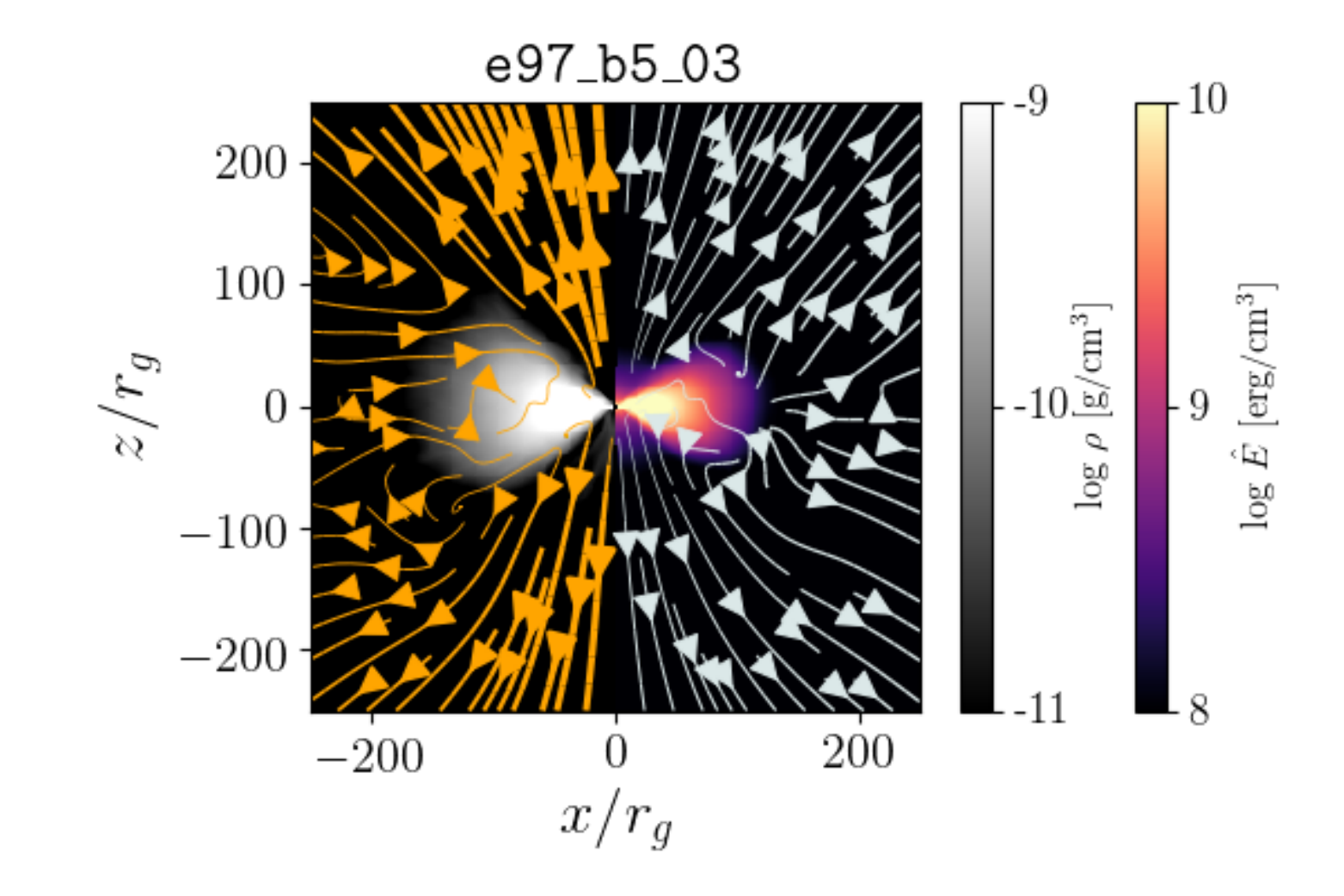}
    \caption{Here we show the mass density (left) and radiation energy density (right) for \texttt{e97\_b5\_01} at $t=40,000\,t_g$ (top panel) and \texttt{e97\_b5\_03} at $t=60,000\,t_g$ (bottom panel). The orange arrows indicate the gas velocity (left) and the white arrows indicate radiative flux (right). We discuss the figures in the text.}
    \label{fig:accRAD}
\end{figure}

As long as the stream is present, new gas with high eccentricity is supplied to the disk and the mean eccentricity remains close to the initial value injected. The self intersection shock leads to significant dissipation and a circularized disk fills radii up to $r<100\, r_g$. There is prompt accretion both through the accretion disk and of material that directly accretes onto the BH at angles above/below the disk. This is indicated for \texttt{e97\_b5\_01} by the velocity vectors directed towards the BH near the pole in Figure \ref{fig:accRAD}.

\begin{figure}
    \centering{}
	\includegraphics[width=0.84\columnwidth]{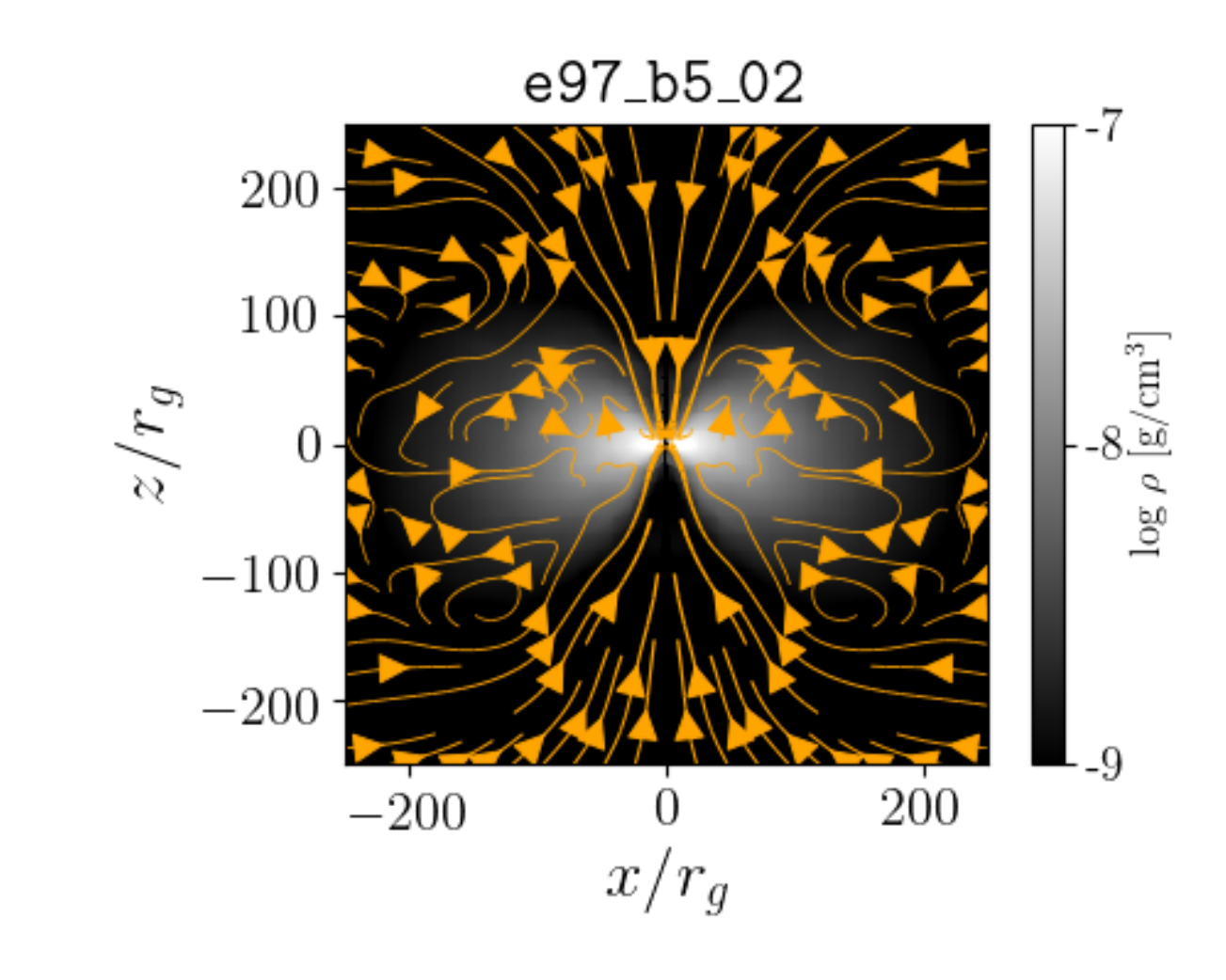}\\
	\includegraphics[width=0.84\columnwidth]{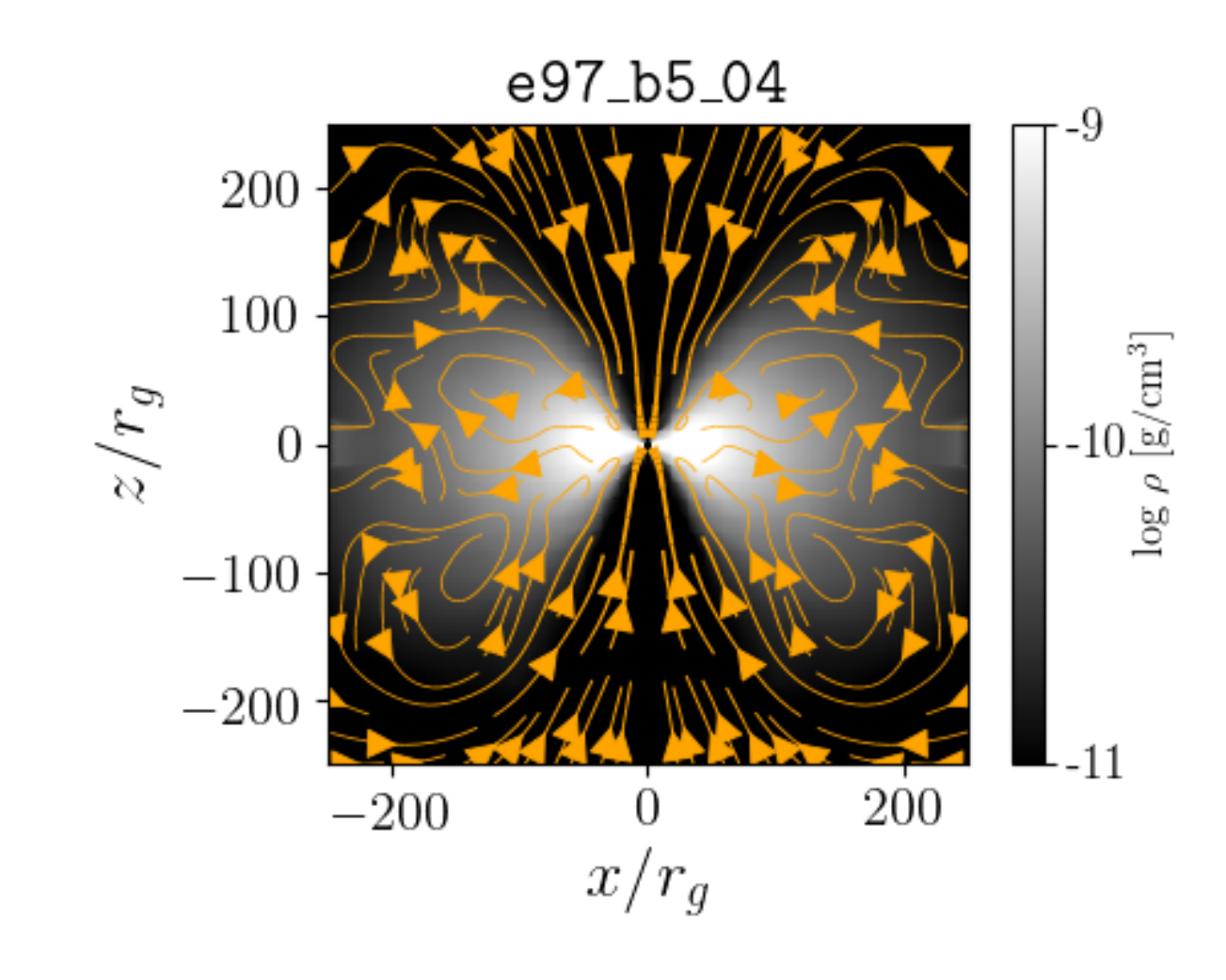}
    \caption{Here we show the mass density (left and right) for \texttt{e97\_b5\_02} at $t=60,000\,t_g$ (top panel) and \texttt{e97\_b5\_05} at $t=60,000\,t_g$ (bottom panel). The orange arrows indicate the gas velocity. We discuss the figures in the text.}
    \label{fig:accHD}
\end{figure}

\begin{figure}
    \centering{}
	\includegraphics[width=\columnwidth]{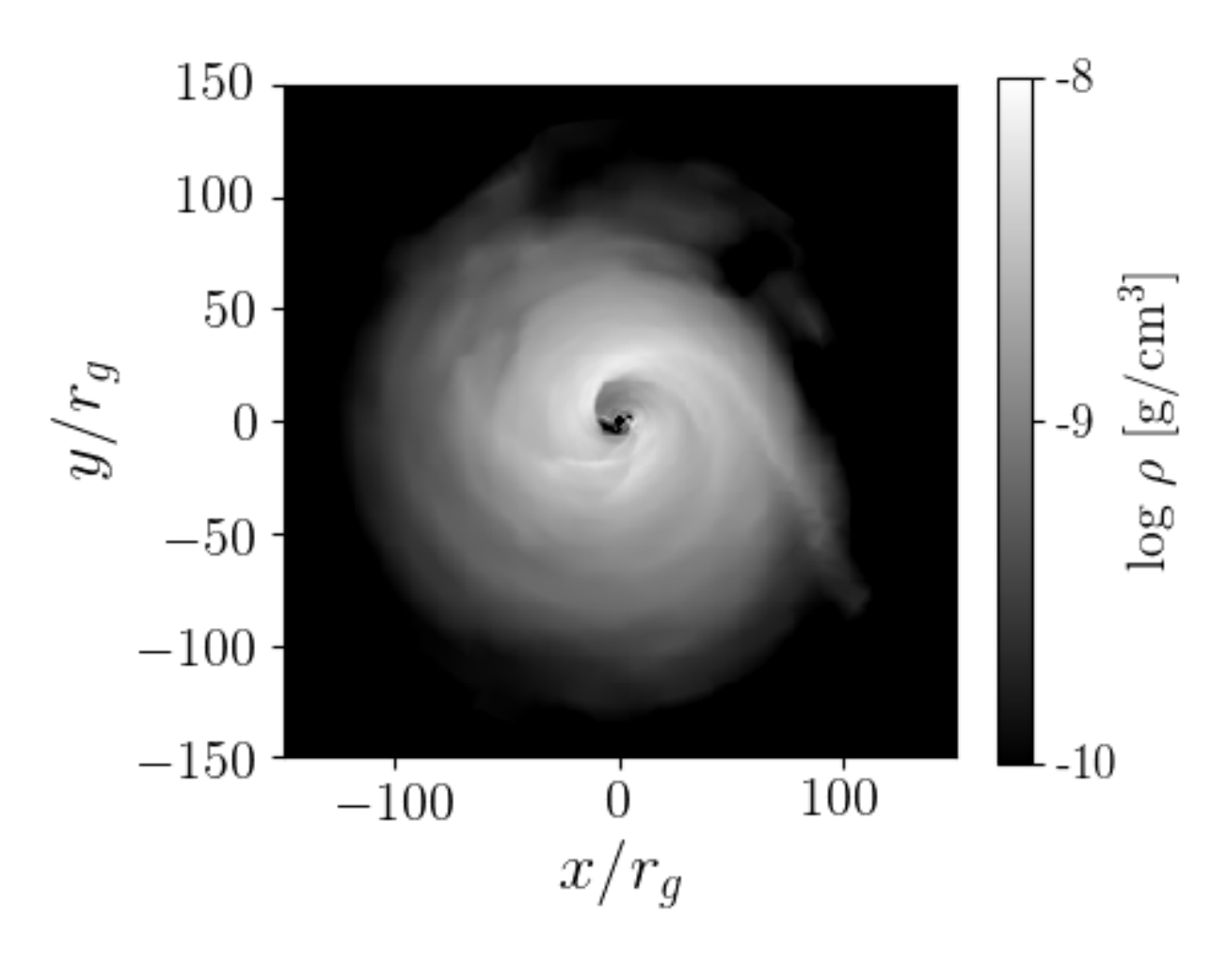}
    \caption{We show an equatorial slice of the gas density at $t=60,000\,t_g$ for \texttt{e97\_b5\_03}. There are spiral density waves present and the disk retains asymmetry owing to its eccentricity.}
    \label{fig:latedisk}
\end{figure}

After the stream injection ends, the already mildly circularized disk material continues to interact and circularize. By the end of each simulation, the disk has stopped evolving in terms of its eccentricity and has settled into a disk of nearly uniform scale height. We show the azimuth averaged vertical structure of the final stage for each simulation in Figures \ref{fig:accRAD} and \ref{fig:accHD}. We also show the equatorial plane for \texttt{e97\_b5\_03} in Figure \ref{fig:latedisk}.

%%%%%%%%
%Begin Figures
%%%%%%%%

\begin{figure}
    \centering{}
	\includegraphics[width=\columnwidth]{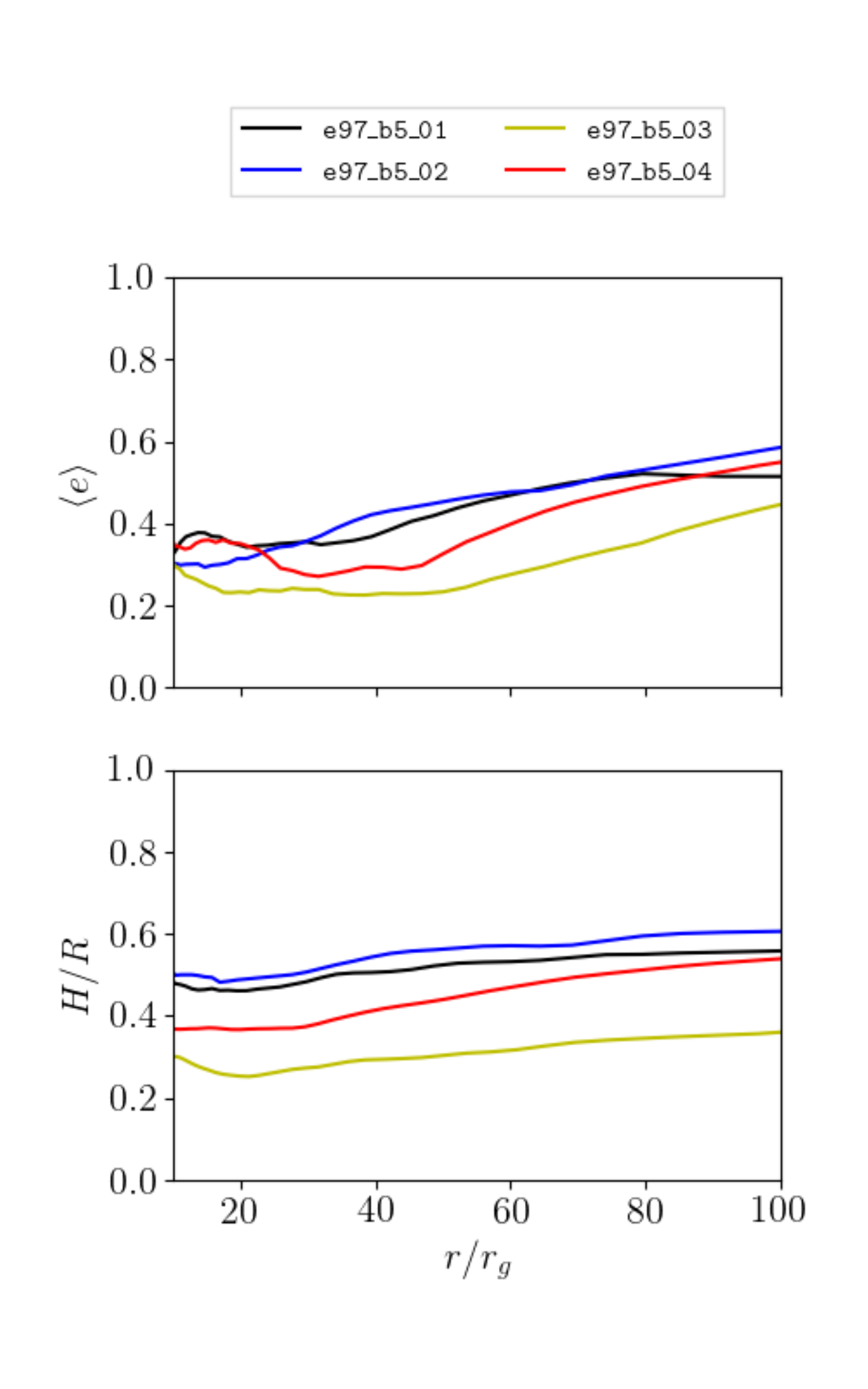}
    \caption{We show the mean eccentricity (top) and the density scale height (bottom) as a function of radius at the end of each of the four $e=0.97$ simulations. The density scale height was computed over $|\theta - \pi/2| \leq \pi/4$.}
    \label{fig:scaleheight}
\end{figure}
%%%%%%%%
%End Figures
%%%%%%%%

\subsection{Accretion Disk Properties}

Here we discuss the end state disk properties. We compute the mean eccentricity within $|\theta - \pi/2| \leq \pi/6$ as in Equation \ref{eq:meanecc}. The circularization is very rapid in each simulation. The majority of the disk mass is near the circularization radius ($R_{\rm{circ}}\equiv2R_p\approx20\,r_g$) and reaches an eccentricity of $e\approx0.3$ in the innermost disk by the end of each simulation. The eccentricity and density scale height of each simulation is shown in Figure \ref{fig:scaleheight}. For \texttt{e97\_b5\_01} and \texttt{e97\_b5\_02}, the results are nearly identical, reflecting the strong radiation trapping which prevents the disk from cooling. For \texttt{e97\_b5\_03} and \texttt{e97\_b5\_04}, the combined effects of a lower accretion rate and radiative cooling result in \texttt{e97\_b5\_03} having a scale height of $H/R\approx 0.25$ while \texttt{e97\_b5\_04} remains quite thick with $H/R\approx 0.4$.

\begin{figure}
    \centering{}
	\includegraphics[width=\columnwidth]{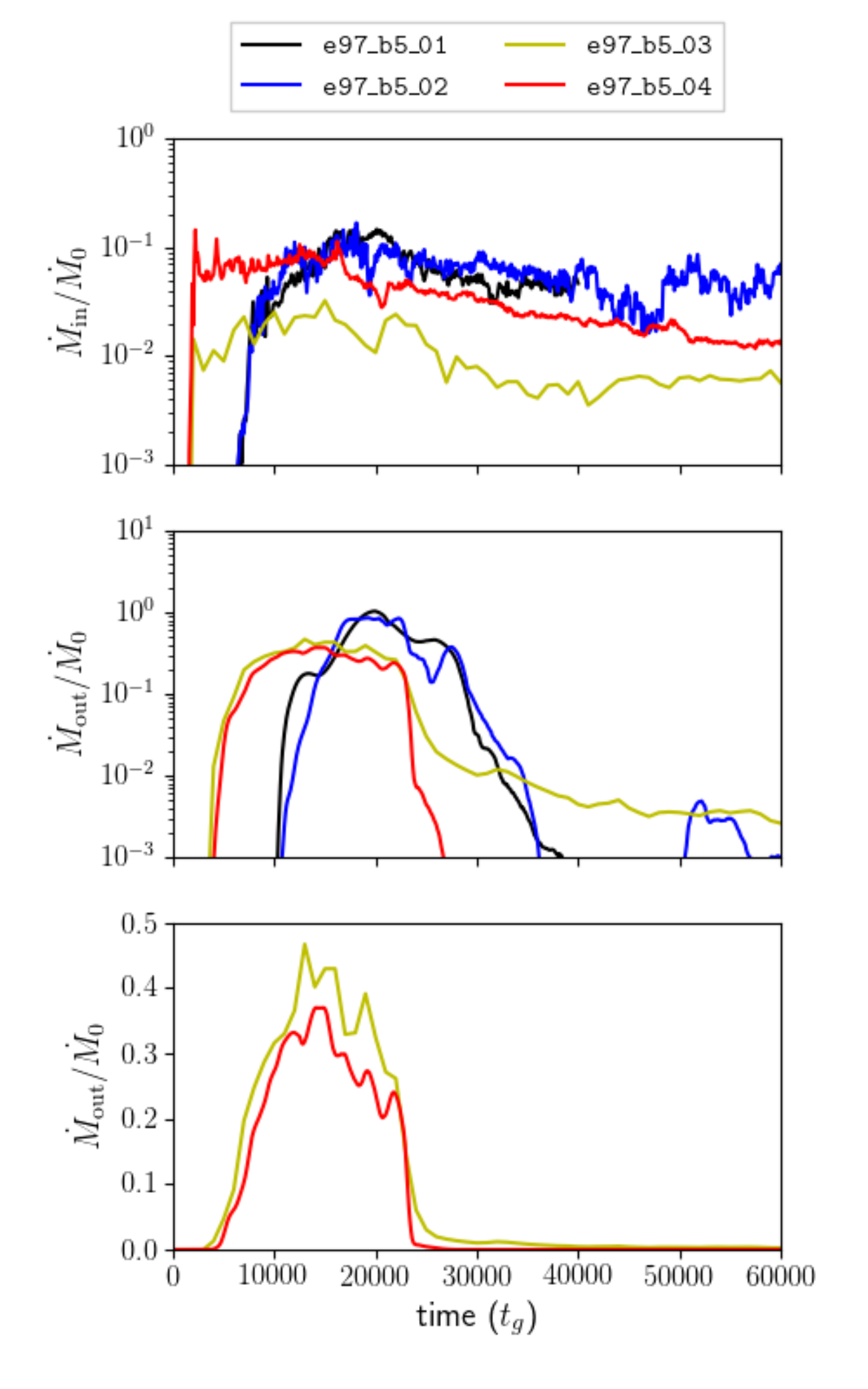}
    \caption{Here we show the rate of inflow of gas crossing the horizon (top) and the rate of ouflow of unbound gas crossing the radius $r=600\, r_g$ (middle). We also show  the mass outflow rate of unbound gas on a linear scale for \texttt{e97\_b5\_03} and \texttt{e97\_b5\_04} (bottom) to better compare these two models.}
    \label{fig:acc}
\end{figure}

The rate of mass inflow through the horizon ($\dot{M}_{\rm in}$) and mass outflow through radius $r=600r_g$ ($\dot{M}_{\rm out}$), both during and after the stream self intersection has ended, are highly super-Eddington (Figure \ref{fig:acc}). We note that for the simulations where more mass is injected (\texttt{e97\_b5\_01} and \texttt{e97\_b5\_02}), there is little difference in the accretion rates. However, the simulations where less mass is injected (\texttt{e97\_b5\_03} and \texttt{e97\_b5\_04}) behave very differently with/without radiation. The accretion rate is nearly an order of magnitude lower when radiation is included. This is a result of the disk and outflow being less optically thick, so the radiation is not as effectively trapped and can diffuse through the disk and outflow. As a result, although the gas is initially more bound, the escaping radiation can accelerate the gas outward and so some of the dissipated orbital energy gets converted back into outflowing kinetic energy. The net effect is that less gas accretes onto the BH and the disk becomes thinner (Figure \ref{fig:scaleheight}),  whereas the outflow is amplified. 

We note that there is a delay in the rise to both peak accretion rate and peak outflow rate in \texttt{e97\_b5\_01} and \texttt{e97\_b5\_02} in comparison to the \texttt{e97\_b5\_03} and \texttt{e97\_b5\_04}. This delay is due to the added BH spin. Such a delay was also observed by \citet{Liptai2019}.

After the self intersection ends, the disk quickly settles into a geometrically thick disk (Figures \ref{fig:accRAD} and \ref{fig:accHD}). This is not surprising since the shock heated gas cannot efficiently cool radiatively due to the dense outflow/inflow which traps radiation. Comparing \texttt{e97\_b5\_03} with \texttt{e97\_b5\_04} shows the effect of radiation quite clearly. The inflow of material near the poles has ended and instead an optically thin funnel forms through which radiation can escape. The gas surrounding the disk has lower density for \texttt{e97\_b5\_03} since the radiation essentially strips the surrounding atmosphere. The radiation has driven a wind which escapes at significantly higher velocities near the poles than near the equatorial plane. We discuss this more in the following subsection.

We confirm that radiative diffusion is relevant in \texttt{e97\_b5\_03} by calculating the time scale for diffusion and advection directly. In optically thick regions, we estimate the ratio between the time scales for diffusion and advection as $t_{\rm{diff}}/t_{\rm{adv}}\approx 3\tau_z (H/R) v$. The opacity is estimated using equation \ref{eq:verttau}.

After the self intersection has ceased, we find that within $r<20 r_g$ the inflow velocity increases towards the BH horizon with a minimum value of $10^{-3}c$ and a maximum of $0.6c$. For $r>20\, r_g$ the inflow velocity is nearly constant with a value of $10^{-3}c$. The vertically integrated optical depth for \texttt{e97\_b5\_03} is $\lesssim 10^3$, but in the case of \texttt{e97\_b5\_01} it is nearly $30$ times larger, reflecting the increased mass injection. As a consequence, for \texttt{e97\_b5\_03} we find that $t_{\rm{diff}}/t_{\rm{adv}} \approx 1.5$ for radii $r>20r_g$ while for $r<20r_g$ the ratio becomes $t_{\rm{diff}}/t_{\rm{adv}} > 1.5$. The similar time scales implies that radiation may diffuse through the disk efficiently. In the case of \texttt{e97\_b5\_01}, $t_{\rm{diff}}/t_{\rm{adv}} \approx 45$ for $r>20\,r_g$ and the ratio only grows at smaller radii so we confirm that diffusion is not an efficient transport mechanism. The significance of advection due to the enhanced opacity can also be seen in Figure \ref{fig:accRAD} as the fluid velocity and radiative flux vectors are generally in the same direction for \texttt{e97\_b5\_01} whereas this is not always the case for \texttt{e97\_b5\_03}. For example, near the edge of the disk in the equatorial plane, the fluid velocity indicates gas is moving towards the BH while the radiative flux indicates radiation is flowing away.

%%%%%%%%
%Begin Figures
%%%%%%%%
%%%%%%%%
%End Figures
%%%%%%%%

\subsection{Outflows}

We show the outflow of unbound gas crossing the shell at $r=600\,r_g$ in Figure \ref{fig:acc}. There are two distinct outflows. The first is launched by the self intersection shock and ends shortly after the least bound part of the stream is injected at $t=14,467\, t_g$. Due to the different injection radii used, the onset of the self intersection outflow begins at $t\approx 1000 t_g$ for $R_{\rm{inj}}=200\,r_g$ and $t\approx 8000 t_g$ for $R_{\rm{inj}}=500\,r_g$. This initial outflow carries an enormous amount of mass and energy. We find that 30-40\% of the injected mass ends up being ejected in an outflow with a similar distribution to \texttt{e99\_b5\_01} in Figure \ref{fig:mdout} at a velocity of $v\approx 0.1c$.

We note that a similar periodicity in the outflow to that described in \S\ref{sec:resultse99} for \texttt{e99\_b5\_01} can be seen in two of the less eccentric models when the mass outflow is highest. For \texttt{e97\_b5\_02}, the period and characteristic radius are slightly larger at $P\approx 4000\, t_g$ and corresponding Keplerian radius of $\approx 75\, r_g$. For \texttt{e97\_b5\_04}, the period and characteristic radius are slightly different however, as we find $P\approx 2000\, t_g$ and corresponding Keplerian radius of $\approx 50\, r_g$. Interestingly, \texttt{e97\_b5\_01} does not show strong periodic behavior in its outflow. We note that the time spacing for data in \texttt{e97\_b5\_03} is $1000\,t_g$, so the periodic behaviour is not well represented in the outflow curve.

The bottom panel in Figure \ref{fig:acc} shows the difference in the outflow during the self intersection shock for \texttt{e97\_b5\_03} and \texttt{e97\_b5\_04}. \texttt{e97\_b5\_04} ejects roughly 30\% of the injected mass while \texttt{e97\_b5\_03} ejects nearly 40\%. This additional outflow is the result of radiation accelerating more gas towards unbound energies in regions where the diffusion timescale is short.

After the disk has become substantially circularized, we find that there is an additional radiation driven outflow in \texttt{e97\_b5\_03}. This outflow is characterized by a lower total mass ejection $\dot{M}_{\rm{out}}\approx3 \dot{M}_{\rm{Edd}}$ and a less uniform velocity distribution (see the bottom panel in Figure \ref{fig:accRAD}). Similar to other super-Eddington accretion disks, a high velocity outflow with $v\sim0.1 c$ is ejected. A significantly slower wind ($v\sim0.01$) is ejected at angles near the equatorial plane. 

While there is occasionally a small net outflow in \texttt{e97\_b5\_01}, \texttt{e97\_b5\_02}, and \texttt{e97\_b5\_04} after the self intersection ends (i.e. the outflow between $t=50,000-55,000\,t_g$ for \texttt{e97\_b5\_02} in the middle panel of Figure \ref{fig:acc}), the specific kinetic energy carried by the outflow is much smaller as the velocity of the outflowing gas is $v<0.01c$.

At present, the computational costs of following the disk evolution well beyond the peak fallback phase (i.e. up to 60 days for a realistic TDE) with radiation and high resolution is substantial. However, if TDEs evolve towards a state where a compact, mild eccentricity disk is embedded in a low optical depth atmosphere, our results suggest that a slower, but wide angle outflow should be expected. 

%%%%%%%%%%%%%%%%%%%%%%%%%%%%% Figures

\begin{figure}
    \centering{}
	\includegraphics[width=\columnwidth]{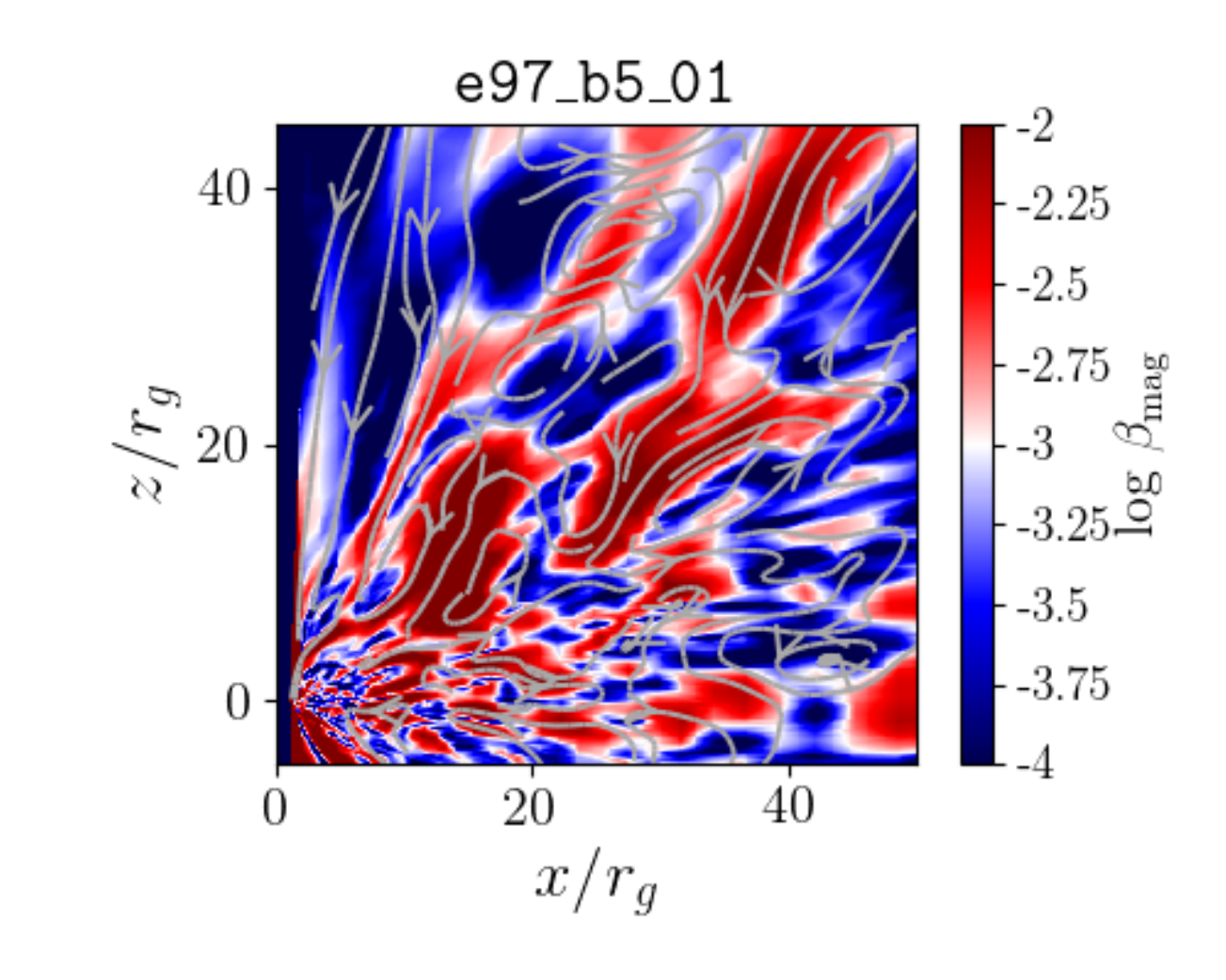}
	\includegraphics[width=\columnwidth]{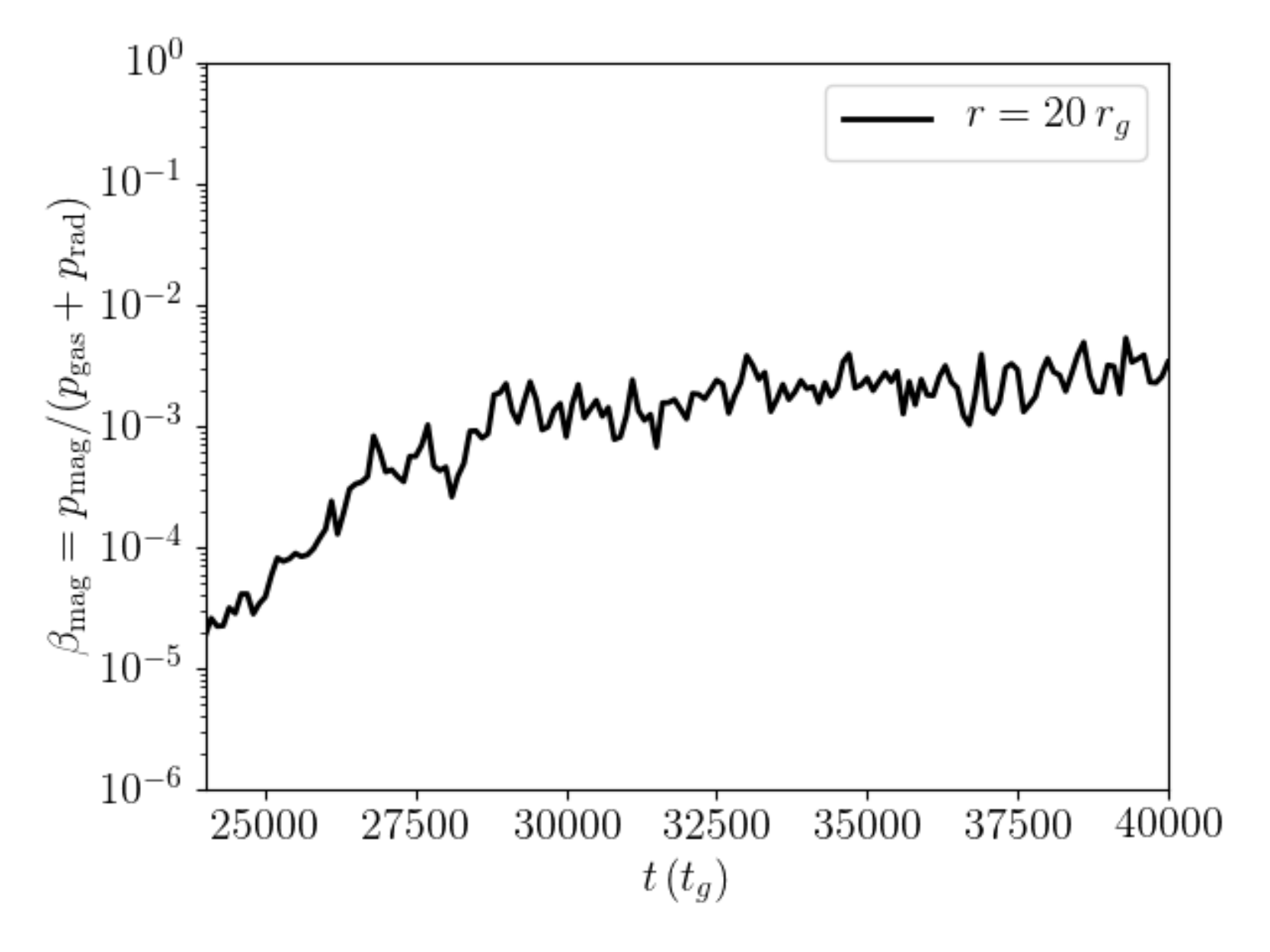}\\
    \caption{In the top panel, we show a snapshot of the magnet pressure ratio (colors) and magnetic field lines (arrows) at the final snapshot ($t=40,000\,t_g$). In the bottom panel, we show the time evolution of the vertically integrated, azimuth averaged magnetic pressure ratio $\beta_{\rm{mag}}$ at $r=20\,r_g$ for times after the self intersection ends ($t>24,000\,t_g$) for \texttt{e97\_b5\_01}. Both panels demonstrate that the magnetic field within the inner disk has grown to $\beta_{\rm{mag}}>10^{-3}$ by the end of the simulation.}
    \label{fig:magbeta}
\end{figure}
%%%%%%%%%%%%%%%%%%%%%%%%%%%%% End Figures

\subsection{The Evolution of the Magnetic Field}

We show the magnetic field properties for \texttt{e97\_b5\_01} in Figure \ref{fig:magbeta}. Although we inject a magnetic field with an initial pressure ratio $\beta_{\rm{mag}}=0.01$, after the gas is heated via shocks, the radiation and gas pressure increase but there is no similar enhancement to the magnetic field. As a result, the magnetic pressure ratio drops substantially initially.

After the self intersection of the stream ends, the differential rotation of the disk winds up the magnetic field, causing the field strength to increase over time. The growth of the field is fastest at smaller radii and over time the magnetic pressure reaches $\beta_{\rm{mag}} > 10^{-3}$ in much of the accretion disk.

We injected a single poloidal loop with the stream. This initialization would lead to a large magnetic flux at the BH horizon if the poloidal loops were advected towards the horizon by the disk. If the magnetic flux threading the disk becomes large enough that the gravitational force is balanced by the outward magnetic pressure, the disk will become magnetically arrested. If this scenario occurs around a rotating BH, a powerful jet could be produced. However, the violent interaction of the stream leads to a disordered magnetic field and a weak magnetic flux at the BH horizon which is well below the limit for a magnetically arrested disk (MAD, \citealt{Narayan2003,Tchekhovskoy2011}).

The magnetic field in a differentially rotating flow is unstable to the MRI \citep{Balbus1991}. The magnetic field strength has not saturated by the end of the simulation and is expected to continue to grow. Local shearing box simulations of the evolution of magnetic fields under Keplerian shear have demonstrated that this growth occurs until $\beta_{\rm{mag}}\approx 0.1$. After the field grows enough for rotational instabilities to be triggered, the induced viscosity is expected to dominate the global dynamics. By the end of the simulation, the MRI quality factors ($Q_\theta$, $Q_\phi$) have reached $\sim 10$, which is the minimum required for MRI resolution, in some regions but only close to the BH ($r<10r_g$). We conclude that for the relatively short time scales simulated in this work, the magnetic field is not expected to change our results. A similarly weak magnetic field growth was found by \citet{Sadowski2016a}. 

We caution that the field strength and polarity chosen in both our work and \citet{Sadowski2016a} is somewhat arbitrary. The primary motivation for the initial field strength is such that the initial dynamics does not become field dominated. There may be exotic scenarios in which the field does become dynamically important. For instance, if a fossil magnetic field gets advected inwards by the stream or if a highly magnetized star which has had its field amplified by multiple partial disruptions becomes fully disrupted, the supplied field may modify the dynamics or produce a jet \citep{Kelley2014,Guillochon2017,Bonnerot2017}.

The initial growth of the poloidal magnetic field is very weak; however, \citet{Liska2020} demonstrate that MHD instabilities can lead to poloidal field being generated from a purely toroidal field. The TDE disk formation process naturally leads to a primarily toroidal field, but if such instabilities exist in TDE disks a strong poloidal component may generate and advect towards the BH, possibly launching a jet. The resolution of the simulations discussed herein is too small compared to the level required to generate a significant poloidal magnetic flux from the toroidal component. \citet{Liska2020} find that up to 15\% of the toroidal flux is converted to a poloidal component. We compute $\Phi_t(r=200\,r_g) = \int_0^{r=200r_g} B^\phi dA_{r,\theta} $, which covers the entire disk. The estimated toroidal flux is $\Phi_t(r=200\,r_g) \approx 30$, which would only lead to a poloidal magnetic flux of $\Phi_p\approx 5$ within the disk. This is a factor of 10 smaller than the flux found in a fully MAD accretion disk \citep{Tchekhovskoy2011}, but we note that our initial magnetic field is simply initialized with $\beta_{\rm{mag}}=0.01$, so we do not claim that this result is representative of more realistic TDE simulations.

%%%%%%%%%%%%%%%%%%%%%%%%%%%%% Figures

\begin{figure}
    \centering{}
	\includegraphics[width=\columnwidth]{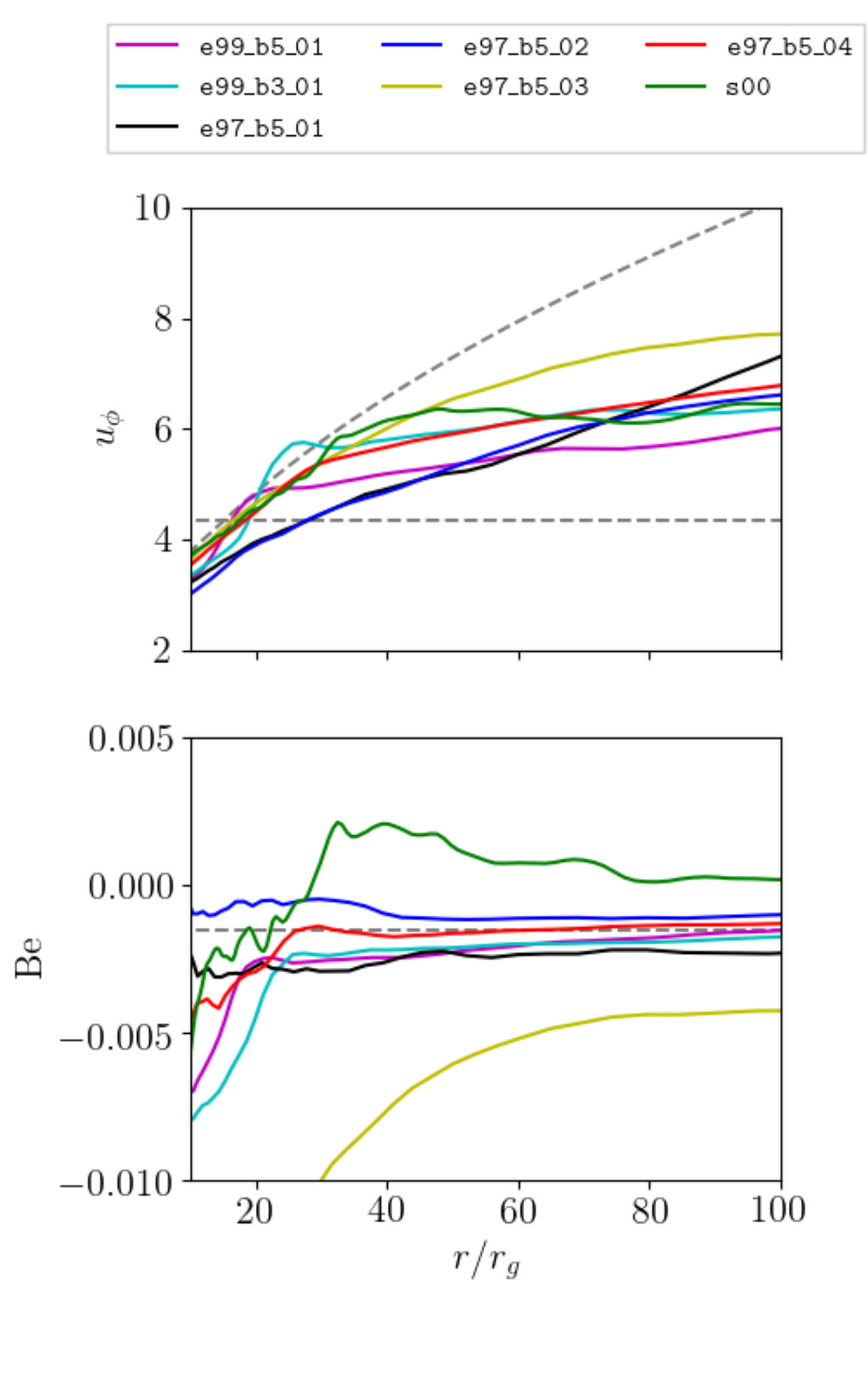}
    \caption{Here we show the angular momentum (top) and Bernoulli (bottom) profiles as a function of radius, averaged over $\pm\pi/4$ from the equatorial plane, at the final epoch of each simulation considered herein. For comparison, we also show the corresponding profiles for the model \texttt{s00} (solid green curves) which was presented in \citet{Curd2019}. In the top panel, we show the Keplerian profile (sloping dashed gray curve) and the angular momentum of the disrupted star with $\beta=5$ (horizontal dashed gray line) for comparison, and in the bottom panel, we show the initial binding energy of the star for $e=0.97$ (horizontal dashed gray line). Note that \texttt{s00} was assumed to have formed from an $e=1$ star so the initial binding energy was close to zero. By the end of that simulation, some of the disk material has become unbound, as expected for accretion flows where the initial gas is nearly unbound \citep{Coughlin2014}.}
    \label{fig:angmom}
\end{figure}

%%%%%%%%%%%%%%%%%%%%%%%%%%%%% End Figures

\section{Discussion} \label{sec:discussion}

\subsection{Comparison with Previous Work}

Here we compare the simulations presented in this work with previous simulations of TDE disks. We differ from similar works in that we inject a TDE stream using theoretical estimates for the binding energy and angular momentum. In addition, ours are the first simulations of a close TDE which evolve both the radiation field and the gas. We have simulated spinning and non-spinning BHs.
In all the simulations, a geometrically thick disk with $H/R \approx 0.25-0.5$ is formed, with significantly circularized gas.

\citet{Sadowski2016a} performed the first general relativistic magnetohydrodynamics (GRMHD) simulation of the TDE disk formation for a $\beta=10, e=0.97$ disruption, but for a $M=10^5\, M_\odot$ BH and a $0.1\, M_\odot$ star. They initialized the TDE by following the disruption of the star using an SPH simulation. As in the present paper, they found that disk circularization occurred rapidly, that the magnetic field had little impact on the overall evolution, and that accretion was driven by hydrodynamic turbulence rather than the MRI. Their estimated peak luminosity was highly super-Eddington ($\sim 40 \, L_{\rm{Edd}}$), whereas our simulations only reach $3-5\,L_{\rm{Edd}}$. This is perhaps owing to the different methodology they used to estimate the emerging radiation. We directly integrate the radiative flux at the photosphere while they estimate the flux using the gas internal energy. Furthermore, the photosphere radius was beyond the outer domain in their work.

As we show in Figure \ref{fig:angmom}, the angular momentum of the simulations we present in this work are sub-Keplerian and have shifted significantly from the injected angular momentum. At smaller radii ($r\lesssim 40$), the angular momentum is much closer to Keplerian, reflecting the relatively low mean eccentricity. Other works that have studied the disk formation of $\beta>1$ disruptions have also found significant evolution of the specific angular momentum \citep{Shiokawa2015,Bonnerot2016,Hayasaki2016}. This behavior is markedly different than the hydrodynamical TDE simulations considered by \citet{Sadowski2016a} who find that the angular momentum of the final disk remains close to that of the initial star. 

In Figure \ref{fig:angmom}, we also compare our results with the GRRMHD simulations of super-Eddington accretion flows presented in \citet{Curd2019}. \citet{Curd2019} assumed an initial torus with constant angular momentum and binding energy, and their model resulted in a nearly constant angular momentum disk that only became Keplerian within regions that had reached inflow equilibrium ($r< 25\, r_g$, Figure \ref{fig:angmom}). Much of the disk in \texttt{s00} at larger radii maintains the initial angular momentum of the star, $l \approx \sqrt{2R_t/\beta}\approx 6$. While the general behavior of the inner accretion flow is similar, our simulations predict that the angular momentum profile of the outer disk does not actually maintain the angular momentum of the disrupted star and instead is shifted to higher angular momenta. This is likely the result of the disk/stream interactions wherein the incoming stream has its momentum driven to larger values via self intersection. 

The Bernoulli number (or specific energy) of the accretion disk decreases slightly when radiation is evolved with the gas. We find that \texttt{e97\_b5\_01}, \texttt{e97\_b5\_02}, and \texttt{e97\_b5\_04} have Bernoulli profiles that remain close to the binding energy of the star; however, \texttt{e97\_b5\_03} has a profile that is generally 2-3 times more negative at large radii than the Bernoulli of the initial star (Figure \ref{fig:angmom}). The reason for this difference is that unlike the other simulations, \texttt{e97\_b5\_03} can efficiently convert its heat energy (which is predominantly radiation) into kinetic energy and drive strong outflows (as we described in \S\ref{sec:resultse97}). As a result, the tenuously bound gas that would otherwise make up its atmosphere gets ejected and a tightly bound disk is left. Both \texttt{e99\_b5\_01} and \texttt{e99\_b3\_01} have a final Bernoulli that is lower than the initial binding energy of the star, but they do not evolve towards a state similar to \texttt{e97\_b5\_03} since the disk is perpetually replenished with gas from the stream which has a low binding energy. 

An interesting aspect of \texttt{s00} from \citet{Curd2019} in comparison to models presented in this work is the positive Bernoulli gas at radii $20\,r_g < r < 80\,r_g$. Model \texttt{s00} was initialized with a large scale torus of marginally bound gas extending to $r=5\times10^3\,r_g$ and the accretion flow deposits energy in the gas at smaller radii. But, instead of simply getting ejected, this gas is prevented from escaping by the material at even larger radii which is pressure supported. In order to drive a substantial wide angle outflow, enough work must be done on the outer component of the torus by the accretion flow/wind to unbind it. Note that \citet{Curd2019} implemented a torus model which more closely resembles the description of TDE accretion disks proposed by \citet{Coughlin2014} whereas \texttt{e97\_b5\_03} appears to form smaller scale disks which resemble typical super-critical accretion disks \citep{Sadowski2014}. More realistic simulations are required to examine whether real TDEs are even capable of forming a torus with bound material at large radii or if the inclusion of radiation leads to a smaller scale disk where the tenuously bound material is largely removed by the time an accretion disk forms.

The outflows launched after the self intersection ends in \texttt{e97\_b5\_03} are wide angle with a significant wind flowing out in all directions (Figure \ref{fig:accRAD}). This is in stark contrast to the model \texttt{s00} in \citet{Curd2019}, which was initialized with a large scale torus, assumed to have formed due to rapid circularization. The torus in \texttt{s00} has a small opening angle funnel which confines the outflow and results in a $v\sim0.1-0.2c$ outflow which only covers an angle of $\sim 12^{\circ}$.

\subsection{Comparison with Other Disk Formation Simulations}

\citet{Shiokawa2015} studied the disk formation following a white dwarf disrupted by a low mass BH ($M_{\rm{BH}} = 500\, M_\odot$) on a parabolic orbit. While they study a different region of parameter space ($\beta=1$) than we do ($\beta=5$, 3), their results are complementary. They identify various shocks due to stream self interaction which dissipate energy throughout the simulation and lead to circularization. Since the self intersection radius in their simulation occurs at a radius of $r\approx1000 r_g$ the kinetic energy dissipated by the initial shock is substantially smaller. The disk remains significantly eccentric after several times the fallback time. 

\citet{Bonnerot2016} performed SPH simulations of stellar disruptions for stars of various impact parameters ($\beta=1-5$) for eccentricities of $e=0.8-0.95$. They find that the disk efficiently circularizes within several times the fallback time and the cooling efficiency has a significant impact on the disk geometry as it circularizes. \citet{Hayasaki2016} performed similar simulations and obtained similar results. They also show that misaligned orbits around spinning BHs lead to less efficient dissipation, but a geometrically thick disk forms as long as the radiative cooling is inefficient. We find a geometrical thickness of $H/R\approx0.4$ for \texttt{e99\_b5\_01} and \texttt{e99\_b3\_01}, which suggests that radiative cooling is not able to dissipate much of the energy deposited in the gas via shocks for $\beta\geq3$ disruptions of near Solar mass stars, at least during the initial disk formation. On the other hand, \texttt{e97\_b5\_03} demonstrates quite clearly that radiative cooling can significantly reduce the disk thickness (Figure \ref{fig:scaleheight}). This was also confirmed by \citet{Bonnerot2021}.

\citet{Liptai2019} investigated the disk circularization for disruptions around spinning BHs in SPH and included general relativistic effects. The disruption parameters they use are very similar to those employed here as the BH mass and stellar mass are identical. The eccentricity is slightly lower however as they choose $e=0.95$. They confirm the general picture of disk formation with regards to the effects of cooling which was first explored by \citet{Bonnerot2016} and \citet{Hayasaki2016}. They find that the dissipation of kinetic energy (or the heating rate due to shocks) for $\beta=5$ disruptions is nearly 2 orders of magnitude higher than for $\beta\sim 1$ disruptions, which suggests that dissipation of kinetic energy increases with increasing impact parameter. Although we do not directly track the dissipation in our simulations, the relative rate of circularization and mass accretion are an indicator of how much kinetic energy is dissipated from the self intersection, thus the dissipation rate in \texttt{e99\_b3\_01} is lower than in \texttt{e99\_b5\_01}. Despite the difference in impact parameter for \texttt{e99\_b5\_01} and \texttt{e99\_b3\_01}, the emerging luminosity and radiation temperature are both generally of the same magnitude. This suggests that at least during the earliest stages of the TDE shortly after the outflow is launched, the emerging luminosity and the dissipation happening beneath the photosphere are not strongly coupled.

\citet{Bonnerot2020} followed the disk formation for a $\beta=1$ disruption using a realistic BH mass and initial stellar eccentricity of $e\sim 1$ using GR SPH. They injected the outflow resulting from self intersection into the domain using a prescription developed by \citet{Lu2020}. Similar to \citet{Shiokawa2015}, they identify complex shocks in the forming disk that dissipate energy and lead to substantial circularization. Dissipation in the secondary shocks in the forming disk outweighed the self-intersection shock by an order of magnitude. They find a heating rate due to shocks that is near the Eddington limit and find that a large fraction of this energy likely participates in the bolometric luminosity since adiabatic losses in their work are expected to be small. In this initial work, the disk structure resembled that of earlier works in that it had substantial geometrical thickness ($H/R\approx1$). 

A follow up study which evolved the radiation with the gas \citep{Bonnerot2021}, showed a disk structure which was substantially thinner while the escaping luminosity was near the Eddington limit. In their simulations, the disk height thinned to $2R_t$ and was nearly constant with radius out to $10R_t$. We similarly find that, when radiative cooling is efficient enough (as in \texttt{e97\_b5\_03)}, the disk thickness decreases, but the disk does not have a constant vertical height in any of the simulations presented in this work. We also note that \citet{Bonnerot2020} and \citet{Bonnerot2021} found that the final disk had an angular momentum sign that was opposite to that of the initial star, but we do not replicate this result even for the least relativistic disruption we simulate ($\beta=3$). 

The radiation temperature of the disk near the BH horizon that \citet{Bonnerot2021} find is much cooler than \texttt{e99\_b5\_01} and \texttt{e99\_b3\_01}. They find a peak radiation temperature of the disk of $T_{\rm{rad}}\approx 10^5$ K whereas we find a peak disk temperature of $T_{\rm{rad}}\approx10^6$. The radiation escaping from the thermalization surface in their work is also cooler as they find typical temperatures of $8\times 10^4$ K and a lower temperature region near the self intersection shock of $3\times10^4$ K, which may lead to observable optical photons. This is in contrast with our typical temperature of $10^5$ K with regions near pericenter reaching as high as $10^6$ K. This large difference is possibly due to the fact that we simulate a higher impact parameter TDE ($\beta=3,5$) wherein the kinetic energy dissipation due to stream self intersection is greater, while \citet{Bonnerot2021} simulate a $\beta=1$ disruption. For instance, \citet{Dai2015} demonstrated that soft X-ray TDEs may be lower mass BHs ($M_{\rm{BH}}<5\times10^6\, M_{\odot}$) disrupting stars on higher impact parameter orbits ($\beta>3$).

\citet{Andalman2020} performed disk formation simulations in which they adopted a BH mass of $10^6\, M_\odot$ and disrupted a solar mass star on a parabolic ($e=1$) orbit with impact parameter $\beta = 7$. They simulate the initial disk formation resulting from the most bound material as the fallback rate of the stream is approaching its peak. They similarly find that a puffy disk which remains highly eccentric forms within a few days. As noted in \S\ref{sec:resultse99}, they identify periodic stream disruptions and posit that it could in principle account for the variability in \textit{Swift} J1644+57. Since the jet power is proportional to the accretion power, if the accretion rate varies during the stream disruption, as it does for some of the stream disruption events in their simulation, this may lead to observable variability. We also identify several stream disruption events in our models \texttt{e99\_b5\_01} and \texttt{e99\_b3\_01}. For \texttt{e99\_b5\_01}, we identify four such events at $t=28,400\, t_g, 32,300\, t_g,39,200\,t_g$ and $50,400\,t_g$, with each disruption lasting roughly $2000\,t_g$.  For \texttt{e99\_b3\_01}, we only identify two such disruptions at $t=7000\,t_g$ and $15,600\,t_g$, with each disruption lasting nearly $3000\,t_g$. There are also weaker interactions during the simulation where the stream is not fully disrupted and instead the angular momentum of the incoming stream is merely pushed to slightly larger values which, as described in \S\ref{sec:e99outflows}, causes variability in the mass outflow. While the accretion rate in Figure \ref{fig:mdote99} is indeed variable, it is not clear that this variability is correlated with the stream disruptions.

Much attention has been given to the general disk formation process in TDEs using hydrodynamic simulations. The present work is only the second to include the magnetic field. Similar to \citet{Sadowski2016a}, we find that the magnetic pressure ratio $p_{\rm mag}/(p_{\rm gas}+p_{\rm rad})$ drops substantially as the stream kinetic energy gets dissipated and the gas/radiation pressure increases. Therefore, the magnetic field has a negligible effect on the evolution of the system. \citet{Bonnerot2021} note that it is possible that the added viscous heating supplied by MRI driven viscocity will play an important role in heating the forming disk, but we leave an exploration of this question to a future study.

\section{Conclusions} \label{sec:conclusions}

We have carried out six simulations of the disk formation following the disruption of a solar mass star on a close orbit around a super massive BH of mass $10^6 \, M_\odot$. We use a novel method of injecting the stream of bound gas into the simulation domain using TDE theory to initialize the inflowing gas. The properties of each simulation are summarized in Table \ref{tab:tab1}. 

For \texttt{e99\_b5\_01} and \texttt{e99\_b3\_01}, the chosen eccentricity to set the orbital dynamics is $e=0.99$, but we set the mass injection rate to that of a parabolic disruption in order to approximate a more realistic TDE. We evolve the radiation with the gas to capture the effects of radiative cooling and radiative diffusion. The impact parameter is $\beta=5$ for \texttt{e99\_b5\_01} and $\beta=3$ for \texttt{e99\_b3\_01}. In both simulations, the injection of the TDE stream is still ongoing by the last epoch. 

We also performed a suite of simulations with $e=0.97$  (\texttt{e97\_b5\_01}, \texttt{e97\_b5\_02}, \texttt{e97\_b5\_03}, and \texttt{e97\_b5\_04}) to validate the injection method that we implement in this work against previous work, and to investigate the impact of radiation during and after disk formation. Each of these four simulations was initialized with the binding energy and angular momentum corresponding to a $\beta=5$, $e=0.97$ disruption. For \texttt{e97\_b5\_01} and \texttt{e97\_b5\_02}, we inject a full solar mass but \texttt{e97\_b5\_01} includes both radiation and a magnetic field whereas \texttt{e97\_b5\_02} was done in pure hydrodynamics. For \texttt{e97\_b5\_03} and \texttt{e97\_b5\_04}, we inject 4\% of a solar mass such that radiative diffusion is possible. Model \texttt{e97\_b5\_03} includes the effects of radiation, while \texttt{e97\_b5\_04} was done in pure hydrodynamics.

We summarize our findings below:

\begin{itemize}
    \item \textit{Disk Formation} - For \texttt{e99\_b5\_01} and \texttt{e99\_b3\_01}, which were initialized with $e=0.99$, we find that within the simulated period of $3.5$ days an accretion disk that is mildly circularized with an eccentricity of $e\approx0.6$ forms. The disk is relatively thick with $H/R\approx 0.4$. As expected, the rate of circularization is slower for lower impact parameter orbits owing to the lower energy dissipation. \citet{Andalman2020} and \citet{Bonnerot2021} similarly find that the disk tends to maintain a rather high eccentricity. Similar to previous simulations of less eccentric disruptions \citep{Bonnerot2016,Hayasaki2016,Sadowski2016a,Liptai2019}, \texttt{e97\_b5\_01}, \texttt{e97\_b5\_02}, \texttt{e97\_b5\_03}, and \texttt{e97\_b5\_04} confirm that, for eccentric TDEs where the stream returns in a finite amount of time, the disk eccentricity is able to reach low eccentricities of $e\approx0.3$ once the stream stops supplying new gas. In addition a geometrically thick disk with $0.25 < H/R < 0.5$ is formed in our simulations. In all six simulations, the final disk maintains the same sign of angular momentum as the initial star, but the final angular momentum is shifted to slightly higher values than that of the initial stellar orbit.
    
    \item \textit{Outflows} - For \texttt{e99\_b5\_01}, nearly half of the injected mass is expelled in an outflow due to the violent self intersection of the stream \citep{Lu2020} as well as an additional outflow from shocks in the forming disk. The majority of the unbound outflow is directed in the opposite direction of pericenter (Figure \ref{fig:mdout}). There is also a significant amount of mass expelled at other angles, but the relative mass flux in this more isotropic outflow is about an order of magnitude less compared to the outflow opposite to pericenter.  For \texttt{e99\_b3\_01}, similar behavior is initially observed at the onset of the outflow; however, the fraction of the outflow that is unbound drops to nearly 5-10\% of the injected mass by the end of the simulation. This is the result of stream deflection driven by a self intersection feedback loop decreasing the kinetic energy of the collision.
    
    \item \textit{Periodicity} - For each simulation, we find periodic behavior in the mass outflow rate during the injection of the stream. This is due to a feedback loop caused by the self intersection driving gas flowing to pericenter towards higher angular momentum orbits which weakens the self intersection. For \texttt{e99\_b5\_01} and \texttt{e99\_b3\_01}, some of these periodic events are accompanied by a complete disruption of the incoming stream. This effect was also seen in the simulation presented by \cite{Andalman2020}. We do not see any events of complete stream disruption in \texttt{e97\_b5\_01}, \texttt{e97\_b5\_02}, \texttt{e97\_b5\_03}, or \texttt{e97\_b5\_04}.
    
    \item \textit{Photosphere and Radiation} - The initial radiation from \texttt{e99\_b5\_01} and  \texttt{e99\_b3\_01} is provided by the expanding outflow with a photosphere that is generally at $R_{\rm{ph}}\approx 1-5\times10^{14}$ cm. Over the period of time that we simulate, the luminosity is mildly super-Eddington and reaches $L_{\rm{bol}}\approx 3-5\,L_{\rm{Edd}}$ by the end of each simulation. The photosphere is highly asymmetric in structure with generally smaller radii near pericenter. The estimated radiation temperature at the photosphere is $T_{\rm{rad}}\lesssim10^5$ K for both \texttt{e99\_b5\_01} and  \texttt{e99\_b3\_01}. This is substantially hotter than any optically identified TDE, which have a blackbody temperature of $T_{\rm{bb}}\sim (1-\rm{several})\times10^4$ K.  The minimum radius of the photosphere is only $\sim 3-6\times 10^{13}$ cm. Furthermore, the radiation leaving the thermalization surface along lines of sight near pericenter has a maximum temperature of $10^6$ K and a typical temperature of $T_{\rm{rad}} \approx 3-5\times 10^5$ K. This may lead to observable soft X-ray emission for suitably oriented observers. We determine a radiative efficiency $\eta\approx 0.009-0.014$, which is smaller than the efficiency $0.057$ expected for a Novikov-Thorne disk. We note that the periodic behavior in the mass outflow rate does not appear to produce an observable periodicity in the escaping luminosity.
    
    \item \textit{Effects of Radiation} - The effects of radiation for extremely dense streams, such as in \texttt{e97\_b5\_01}, is negligible. In this case, the radiatiive diffusion time scale is significantly longer than the accretion time scale. Consequently, radiation cannot effectively act on the gas and is simply advected with the gas. In cases where the stream is less dense to resemble realistic TDEs, such as in \texttt{e99\_b5\_01},\texttt{e99\_b3\_01}, and \texttt{e97\_b5\_03}, radiation has a significant impact. The mass outflow launched during the self intersection is amplified due to the acceleration provided by radiation. In addition, the mass accretion rate is significantly less, which leads to a thinner accretion disk. Force supplied via radiation accelerates the gas and essentially redistributes kinetic energy that has been deposited in radiation back into gas kinetic energy. This also explains why the Bernoulli of the disk is lower overall when radiation is included (\texttt{e97\_b5\_03}) when compared to the pure hydrodynamics simulation (\texttt{e97\_b5\_04}). The radiation expels the tenuously bound gas, leaving predominantly more bound gas in an accretion disk with a small scale height.
\end{itemize}

The method of injecting the TDE stream that we have used saves computational resources by skipping the initial disruption stage. By simply modifying the binding energy distribution and angular momentum of the injected gas, as well as the BH properties, this method opens up the potential to study a large parameter space. One caveat in this study is that we are limited to injecting streams with artificially larger initial scale height due to the resolution in the azimuthal direction. Nonetheless, the qualitative agreement of our results with previous work demonstrates that this method accurately reproduces key physical features of TDE evolution. We plan to incorporate the effects of bound-free absorption and obtain detailed spectra from ray traced images with future simulations. We also plan to apply the methods used here to study the early disk formation and radiative properties for other BH masses.

\section*{Acknowledgements}

We are grateful to Ramesh Narayan for many useful comments, discussions, and suggestions throughout this work. We thank Wenbin Lu and Cl\'{e}ment Bonnerot for comments and discussions regarding this work. We also thank Edo Berger, Lars Hernquist, Josh Grindlay for their insight. This work was supported in part by NSF grant AST-1816420, and made use of computational support from NSF via XSEDE resources (grant TG-AST080026N). This work was carried out at the Black Hole Initiative at Harvard University, which is supported by grants from the John Templeton Foundation and the Gordon and Betty Moore Foundation.

\section*{Data Availability}
The data underlying this article will be shared on reasonable request to the corresponding author.

%%%%%%%%%%%%%%%%%%%%%%%%%%%%%%%%%%%%%%%%%%%%%%%%%%

%%%%%%%%%%%%%%%%%%%% REFERENCES %%%%%%%%%%%%%%%%%%

%%%%%%%%%%%%%%%%%%%%%%%%%%%%%%%%%%%%%%%%%%%%%%%%%%

%%%%%%%%%%%%%%%%% APPENDICES %%%%%%%%%%%%%%%%%%%%%

\appendix

%%%%%%%%%%%%%%%%%%%%%%%%%%%%%%%%%%%%%%%%%%%%%%%%%%

\bsp	% typesetting comment
\label{lastpage}
\end{document}